\documentclass[usenatbib]{mn2e}
\usepackage{natbib}
\usepackage{graphicx}
\usepackage{amssymb}
\usepackage{xspace}
\usepackage{epsfig}
\usepackage{times}

\newcommand{\refjnl}[1]{{\em #1\ }}
\newcommand\aj{\refjnl{Astron J.}}%
\newcommand\apj{\refjnl{Astrophys. J.}}%
\newcommand\aap{\refjnl{Astron. Astrophys.}}%
\newcommand\mnras{\refjnl{Mon. Not. R. Astron. Soc.}}%
\title[Temperature in H~{\sc ii} regions] {The effects of spatially 
distributed ionisation sources on
   the temperature structure of H~{\sc ii} regions}
\author[Ercolano, Bastian, Stasi\'{n}ska]{B. Ercolano$^{1,2}$, N. Bastian$^2$,
   G. Stasi\'{n}ska$^3$\\
$^1$Harvard-Smithsonian Centre for Astrophysics, 60 Garden Street, 
Cambridge, MA 02138, USA\\
$^2$Department of Physics and Astronomy, University College London, 
Gower Street, London WC1E~6BT, UK\\
$^3$LUTH, Observatoire de Paris, CNRS, Universit\'e
Paris Diderot ; Place Jules Janssen 92190 Meudon,
France\\
\\}
\date{Received:}

\begin{document}
\maketitle

\begin{abstract}

Spatially resolved studies of star forming regions show that the 
assumption of spherical geometry is not realistic in most cases, with
a major complication posed by the gas being ionised by multiple 
non-centrally located stars or star clusters.
Geometrical effects including the spatial configuration of ionising 
sources affect the temperature and ionisation structure of these 
regions. We try to isolate the effects of multiple non-centrally
located stars, via the 
construction of 3D photoionisation models using the 3D Monte Carlo 
photoionisation code {\sc mocassin} with very simple gas density
distributions, but various spatial 
configurations for the ionisation sources. Our first aim is to study the 
resulting temperature structure of the gas and investigate the 
behaviour of temperature fluctuations within the ionised region. We
show that geometry affects the temperature structures in our models
differently according to metallicity. For 
the geometries and stellar populations considered in our study, 
at intermediate and high metallicities, models with ionising sources 
distributed in the full volume, whose Str\"omgren spheres rarely
overlap, 
show {\it smaller} temperature fluctuation than their central ionisation counterparts, with fully 
overlapping concentric Str\"omgren spheres. The reverse is true at low
metallicities. Finally the true temperature {\it fluctuations} due to
the stellar distribution (as opposed to the large-scale temperature
gradients due to other gas properties) are small in all cases and not
a significant cause of error in metallicity studies. 

  Emission line spectra from H~{\sc ii} regions are often used to 
study the metallicity of star-forming regions, as well as providing a
constraint for temperatures and luminosities of the ionising sources. Empirical 
metallicity diagnostics must often 
be calibrated with the aid of photoionisation models. However, most studies so 
far have been carried out by assuming spherical or plane-parallel 
geometries, with major limitations on allowed gas and dust density 
distributions and with the spatial distribution of multiple, 
non-centrally located ionising sources not being accounted for. We 
compare integrated emission line spectra from our models and quantify 
any systematic errors caused by the simplifying 
assumption of a single, central location for all ionising sources.  
We find that the dependence of the metallicity indicators on the ionisation
parameter causes a clear bias, due to the fact that models with a fully
distributed configuration of stars always display lower ionisation
parameters than their fully concentrated counterparts. The errors found
imply that the geometrical distribution of ionisation sources 
may partly account for the large scatter in
metallicities derived using model-calibrated empirical methods.

\end{abstract}

\begin{keywords}
(ISM): H~{\sc ii} regions; galaxies: abundances; radiative transfer
\end{keywords}
\nokeywords


\section{Introduction}

The ability to measure accurate chemical
abundances in H~{\sc ii} regions in our own and other galaxies is
vital for our understanding of their chemical evolution. Emission 
lines emitted by the nebular gas photoionised by
massive stars provide us with powerful metallicity indicators for near
and intermediate redshift galaxies. The major complication is posed by
the critical task of determining the physical parameters of the
nebula, in particular the electron temperature, $T_{\rm e}$. Failure
to achieve realistic estimates may lead to gross errors in the final
abundance determination. Temperature fluctuations within the nebula 
may be an important cause of
error. Collisionally excited lines (CELs),
which are routinely used in abundance studies, are naturally weighted towards
hotter regions and may therefore lead to underestimating of
the real abundances. Recombination lines (RLs) are less affected by errors
in temperature determinations, and should in theory yield more
accurate results. One of the outstanding problems in nebular
astrophysics is the discrepancy between abundances and electron
temperature estimates obtained from CELs or RLs in Planetary Nebulae (PNe)
(e.g. Rubin et al. 2002; Wesson, Liu \& Barlow 2005; Ercolano et al. 2005;
Liu et al. 2006; Peimbert \& Peimbert 2006 and references therein),
galactic and Magellanic H~{\sc ii} regions (e.g. Peimbert, Peimbert \& Ruiz
2000; Peimbert 2003; Tsamis et al. 2003; Esteban et al. 2004;
Garc{\'{\i}}a-Rojas et al. 2005, 2006) and extragalactic H~{\sc
  ii} regions (e.g. Peimbert \& Peimbert 2003; Peimbert, Peimbert \& Ruiz
2005).  
The cause of this discrepancy is still uncertain; temperature
fluctuations and chemical
inhomogeneities in the gas have both been
 advocated to explain the discrepancy. Photoionisation models including
chemical inhomogeneities have been successful in matching the observed
CEL and RL spectra of some PNe (e.g. Ercolano et al. 2003b) and H~{\sc
  ii} regions (Tsamis \& P\'equignot 2005). However the discrepancy
between the abundances derived from recombination lines (RL) and those 
derived from CELs has been also been attributed to temperature
fluctuations {\it not} caused by abundance inhomogeneities (e.g. Garc{\'{\i}}a-Rojas et al. 2004). 
 The $t^2$ parameter (Peimbert, 1967; Peimbert \& Costero 1969;
 Peimbert 1971), introduced in order to quantify
such fluctuations, can be derived empirically by the comparison of 
temperatures obtained from the depth of the Balmer Jump to those obtained from CELs.
Although the situation for H~{\sc ii} regions is notoriously less worrying 
than for PNe,
the problem still remains that empirically determined $t^2$ values 
are larger than
those indicated by chemically homogeneous photoionisation models, 
indicating that the causes of the proposed temperature fluctuations
are still not understood.

Aside from temperature fluctuations the task of obtaining reliable 
abundance estimates from
CEL spectra of H~{\sc ii} regions is further complicated by 
the fact that often the
complete set of emission lines needed for a direct measurement of the
electron temperatures is not available for all sources; the 4363~{\AA} line of
$[$O~{\sc iii}$]$ and the 5755~{\AA} line of $[$N~{\sc ii}$]$ are
weak and therefore not detected if the spectra do not have good 
signal-to-noise or if the metallicities are high, implying low 
electron temperatures.
A large effort has been made to provide metallicity and
temperature indicators based on various combinations of strong lines
(e.g. Pagel et al. 1979; Alloin et 
al. 1979;  Mc Gaugh 1991; Storchi-Bergmann et al. 1994; V\'{i}lchez \&
Esteban 1996; Van 
Zee et al. 1998;
Pilyugin 2001a; Pettini \& Pagel 2004; P\'erez-Montero \& 
D\'{i}az 2005). H\"agele et al. (2006) have recently proposed a new
methodology for the production and calibration of empirical
relations between the different line temperatures based on 
observational data alone. However the data sets
available are still too limited to provide usable indicators. Currently,
 calibrations of metallicity indicators and ionic temperature relations
 are generally not based on observational data alone, but make use of grids of
photoionisation models, run for a range of metallicities and
ionisation parameters (e.g. McGaugh 1991, Charlot \& Longhetti  2001, 
Kewley \& Dopita 2002). Given the vast parameter space generally 
under investigation,
most studies so far have been carried out with spherically symmetric
or plane parallel geometries, with
major limitations on the allowed density distribution and with the
spatial distribution of the ionising sources not being investigated. 
Nearby H~{\sc ii} regions, however, show complex 
structures
in the distribution of gas and stars, which are often intermixed.

In this paper we carry out a theoretical investigation of the
importance of the effects due to the spatial distribution of the
ionising sources via the construction of a number of 3D
photoionisation models, using the {\sc mocassin} code (Ercolano et
al. 2003a, 2005) for simple gas density distributions and three spatial
configurations for the ionisation sources. 

Our modelling strategy is described in Section~2.
Temperature fluctuations are estimated for our models by computing 
theoretical $t^2$ and $T_0$ values (Peimbert, 1967) for each of our 
models and the results are given and discussed in Section~3. In this 
Section we also compare integrated emission line spectra from such 
configurations to search for systematic errors which may be caused by 
the simplifying assumption of a single, central location for all 
ionising sources. We also test the robustness of a number of 
commonly used ionic temperature relations. A summary of our main
results is presented in Section~4.

\section{Modelling strategy}

\begin{table}
\begin{center}
\caption{Characterising model parameters. $L_{\rm bol}$ is the total
  bolometric luminosity of all stars included. $R_{s}$ is the
  Str\"omgren radius in the case when all stars are located at the
  centre of the sphere/shell. The total number of ionising
  photons and the hydrogen number density are constant for all models
  and are, respectively,  $Q_{\rm H^0}$~=~
  3.80$\cdot$10$^{50}$~sec$^{-1}$ and $N_{\rm H}$~=~100~cm$^{-3}$. 
All shell models have an inner radius of 2.8$\cdot$10$^{19}$~cm.}
\begin{tabular}{cccc}
\hline\noalign{\smallskip}
Z/Z$_{\odot}$ &  $L_{\rm bol}$     & \multicolumn{2}{c}{$R_{s}$} \\  
              &  $[$E40 erg/sec$]$   & \multicolumn{2}{c}{$[$E19 cm$]$}   \\
              &                    & sphere & shell                \\
\hline\noalign{\smallskip}
2.0           &  3.30      &   2.65 &  3.45   \\
1.0           &  3.00      &   2.85 &  3.55   \\
0.4           &  2.64      &   3.20 &  3.75   \\
0.2           &  2.30      &   3.45 &  3.90   \\
0.05          &  1.98      &   3.75 &  4.20   \\
\hline\noalign{\smallskip}
\end{tabular}
\label{tab:mods1}
\end{center}
\end{table}

\begin{table*}
\begin{center}
\caption{Model grid legend. H number density, N$_H$~=~100~$[cm^{-3}]$
  for all models. Figures in this paper rely on colour, please refer
  to on-line version. For black and white copies black is black, blue is dark-gray,
red is medium gray and green is light gray.}
\begin{tabular}{lcccclcccc}
\hline\noalign{\smallskip}
model & star distrib. & geometry  & Z/Z$_{\odot}$ & symbol & model & star distrib. & geometry  & Z/Z$_{\odot}$ & symbol \\
\noalign{\smallskip}                                                    \noalign{\smallskip}	 	 					  	  
CSp2.0 & central       & sphere  & 2.0   & small  magenta circle &CSh0.4 & central       & shell   & 0.4   & small   red square    \\
HSp2.0 & half distrib  & sphere  & 2.0   & medium  magenta circle &HSh0.4 & half distrib   & shell   & 0.4   & medium   red square    \\
FSp2.0 & fully distrib & sphere  & 2.0   & large  magenta circle &FSh0.4 & fully distrib & shell   & 0.4   & large   red square    \\
\noalign{\smallskip}	       	 	       	   	  	\noalign{\smallskip}	         		     	          
CSh2.0 & central       & shell   & 2.0   & small  magenta square &CSp0.2 & central       & sphere  & 0.2   & small   green circle  \\
HSh2.0 & half distrib   & shell   & 2.0   & medium  magenta square &HSp0.2 & half distrib  & sphere  & 0.2   & medium   green circle  \\
FSh2.0 & fully distrib & shell   & 2.0   & large  magenta square &FSp0.2 & fully distrib & sphere  & 0.2   & large   green circle  \\
\noalign{\smallskip}	       	 	    	  			\noalign{\smallskip}			     		          
CSp1.0 & central       & sphere  & 1.0   & small  black circle   &CSh0.2 & central       & shell   & 0.2   & small   green square  \\
HSp1.0 & half distrib  & sphere  & 1.0   & medium  black circle   &HSh0.2 & half distrib   & shell   & 0.2   & medium   green square  \\
FSp1.0 & fully distrib & sphere  & 1.0   & large  black circle   &FSh0.2 & fully distrib & shell   & 0.2   & large   green square  \\
\noalign{\smallskip}	       	 	       	   	  	\noalign{\smallskip}								       
CSh1.0 & central       & shell   & 1.0   & small  black square   &CSp0.05 & central       & sphere  & 0.05   & small   green circle\\
HSh1.0 & half distrib   & shell   & 1.0   & medium  black square   &HSp0.05 & half distrib  & sphere  & 0.05   & medium   green circle\\
FSh1.0 & fully distrib & shell   & 1.0   & large  black square   &FSp0.05 & fully distrib & sphere  & 0.05   & large   green circle\\
\noalign{\smallskip}	       	 	    	  			\noalign{\smallskip}			       		          
CSp0.4 & central       & sphere  & 0.4   & small  red circle     &CSh0.05 & central       & shell   & 0.05   & small   green square\\   
HSp0.4 & half distrib  & sphere  & 0.4   & medium  red circle     &HSh0.05 & half distrib   & shell   & 0.05   & medium   green square\\   
FSp0.4 & fully distrib & sphere  & 0.4   & large  red circle     &FSh0.05 & fully distrib & shell   & 0.05   & large   green square\\   
\hline										
\noalign{\smallskip}	       		    		
\end{tabular}
\label{tab:mods}
\end{center}
\end{table*}


\begin{figure}
\begin{center}
\begin{minipage}[t]{4.cm}
\includegraphics[width=4.cm]{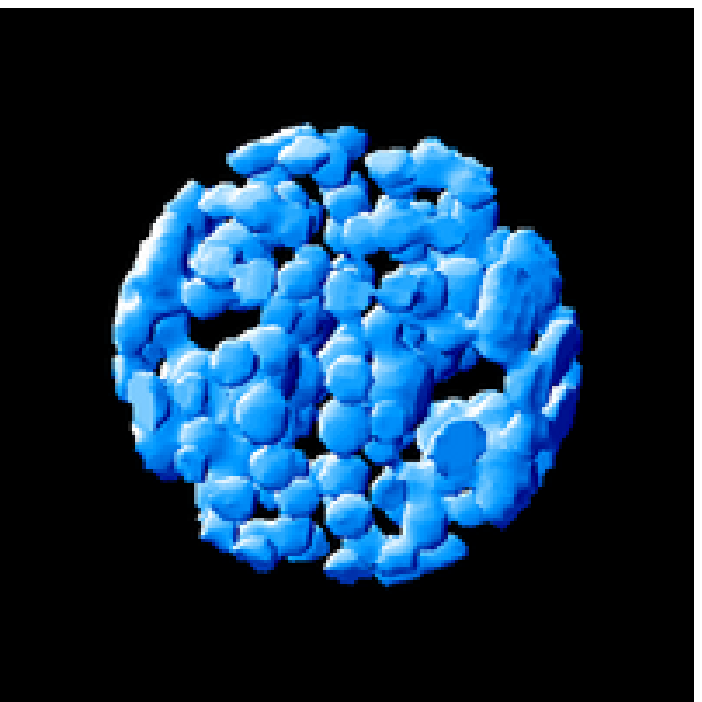}
\end{minipage}
\begin{minipage}[t]{4.cm}
\includegraphics[width=4.cm]{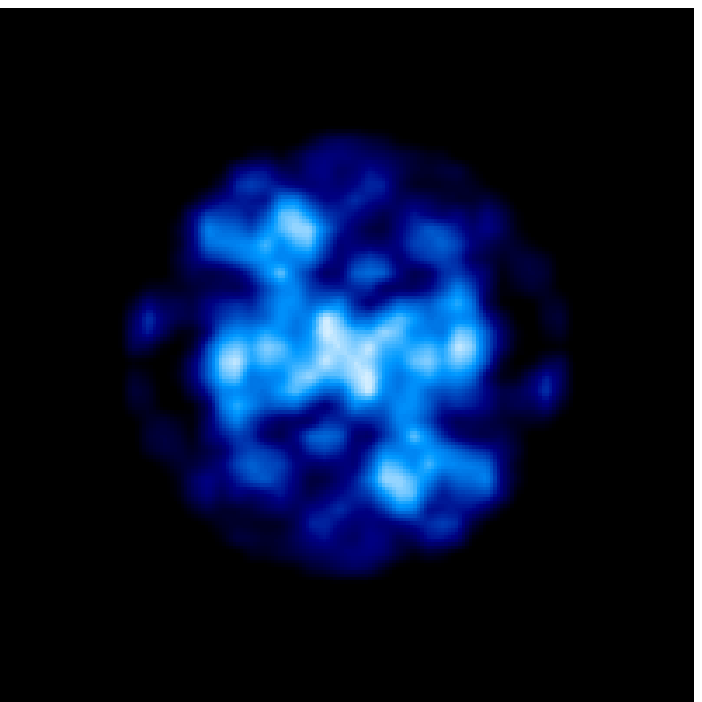}
\end{minipage}
\caption[]{The left panel shows a 3D representation of the Str\"omgren sphere distribution for case F,
plotted as the iso-surfaces where the ionisation fraction of hydrogen
is 0.95. 
The adjacent right panel shows an average projection map of the ionic abundance of H$^{+}$. }
\label{fig:dist}
\end{center}
\end{figure}


The aim of the current investigation is to uncover possible
{\it systematic} differences on the temperature structure and emission 
line spectra of nebulae ionised by a centrally concentrated set of 
stars or clusters versus those ionised by the same set of sources 
which are randomly distributed within the half or full volume, whose 
respective Str\"omgren spheres do or do not overlap. In particular we 
want to test which of the stellar configurations considered produces 
the largest temperature fluctuations and of what magnitude. At this 
stage we do not attempt to assess the absolute strengths and 
weaknesses of one metallicity indicator or ionic temperature relation 
over others, a task that may  only be carried out by a systematic 
investigation of the large parameter space.  In an attempt to isolate
the effects of the stellar distribution from those due to the gas
density distribution, we consider two extremely simple
 geometries - a homogeneous spherical volume and
a homogeneous spherical shell, both of constant hydrogen number
density, $N_{\rm H}$~=~100~cm$^{-3}$. The shell models have inner radii
of 2.8$\cdot$10$^{19}$~cm and Str\"omgren radii (corresponding to the
case when all stars are located at the centre) as listed in
Table~\ref{tab:mods1}. The total number of ionising
  photons is constant for all models
  and is $Q_{\rm H^0}$~=~3.80$\cdot$10$^{50}$~sec$^{-1}$.

For each density distribution we are interested in a comparison between the 
centrally concentrated and the half and fully distributed source cases.
In the remainder of this paper we will refer to models ionised by a
central concentration of stars as models C, those ionised by the same
set of stars distributed in the half and full volumes will be referred
to as models H and F, respectively. Case C is equivalent to models
ionised by a single central source with a total bolometric luminosity
equal to the sum of the bolometric luminosities of the individual
sources (as given in Table~\ref{tab:mods1}) and a spectral shape given 
by the superposition of the spectra of all sources. These models could 
also be performed using a 1D code, with a spherically symmetric gas
density distribution. Similarly a 1D code could also be employed for
case F models where the Str\"omgren spheres do not generally
overlap\footnote{Due to the stochastic nature of the ionisation source
   distribution, it is possible that a small fraction of them may actually
have partially overlapping
   Str\"omgren spheres.}; the left panel of Figure~\ref{fig:dist} shows a 3D
representation of the Str\"omgren sphere distribution for case F,
plotted as iso-surfaces where the ionisation fraction of hydrogen is
0.95. The adjacent panel shows a projected map of the ionic abundance
of H$^{+}$. 
In the intermediate case, case H, the Str\"omgren spheres of
most H~{\sc ii} regions partially overlap; the determination of the radiation
field in the overlap region is not possible without the application 
of a 3D code. As we are interested in a comparison of the three cases
(C, H and F), self-consistency is crucial and for this reason we have
run all models with the same 3D code, {\sc mocassin}. 

Our models include five metallicities ($Z/Z_{\odot}$~=~0.05, 0.2, 
0.4, 1.0, 2.0). The "solar" abundance set uses the values from
Grevesse \& Sauval (1998) with the exception of C, N, O abundances
which are taken from Allende Prieto et al 2002, Holweger 2001 and
Allende Prieto et al 2001, respectively. These were
  scaled to lower and higher metallicities 
considering the empirical abundance trends observed in H~{\sc ii} 
regions by Izotov et al. (2006). 
  In order to maximise the effects of spatially distributed sources
  whose Str\"omgren spheres may overlap (completely or partially), or be
  totally independent, we consider the limiting case of an ionising set
  composed of two stellar populations, a 37~M$_{\odot}$ and a
  56~M$_{\odot}$ population, with half of the total ionising photon
  output [s$^{-1}$] being emitted by each population. These two stellar masses
  were chosen as they have very different $Q_{\rm He^0}$/$Q_{\rm H^0}$
  ratios, where $Q_{\rm H^0}$ is the total output  [s$^{-1}$] of
  H-ionising photons (energy~$>$~1~ryd)  and $Q_{\rm He^0}$ is the total output  [s$^{-1}$]
  of He-ionising photons (energy~$>$~1.8~ryd), and are likely to produce the largest effects
  on the temperature structure and sharpness of the ionisation
  front. The ionising spectra for single-mass stars were computed with 
the {\sc starburst99} spectral synthesis code Leitherer et al. (1999) 
with the up-to-date non-LTE  stellar atmospheres implemented by Smith, Norris \& Crowther
 (2002), using single isochrones for the appropriate stellar 
masses. The models were calculated at metallicities consistent with 
the nebular gas and were obtained for an instantaneous burst, at an 
age of 1 Myr. At solar metallicities for the 37~M$_{\odot}$ stars, 
the stellar atmosphere models emit 32.3\% of luminosity in the 
H-ionising continuum and 7.7\% in the He-ionising continuum, while 
for the 56~M$_{\odot}$ stars these percentages are 47.9\% and 13.7\%, 
respectively. The exact percentages vary with stellar
metallicity, nevertheless the values above are given as a guide to
appreciate the different spectral hardness of the two populations. 

Some other defining parameters of our models, together with their 
nomenclature and associated symbols are given in 
Table~\ref{tab:mods}.

\subsection{The 3D photoionisation code: {\sc mocassin}}
The 3D photoionisation code {\sc mocassin} (Ercolano et al. 2003a, 2005) uses a
Monte Carlo approach to the radiation transport problem and it is
therefore completely independent of geometry and density
distribution. Both the stellar and diffuse components of the radiation
field are treated self-consistently, without the need of
approximations. Multiple ionisation sources can be located at
arbitrary positions in the simulation grid with the only limit being
imposed by computing resources. The atomic data used is frequently
updated and include sets of energy levels, collision strengths and
 transition probabilities from Version 5 of the Chianti database
 (Landi et al., 2005) and the improved H~{\sc i}, He~{\sc i} and He~{\sc ii} 
free-bound continuous emission data recently published by Ercolano \&
Storey (2006). A public version of the {\sc fortran
   90} code, which is fully parallelised using the Message Passing
Interface (MPI) libraries, can be obtained from 
B. Ercolano. Version 2.02.38 was used for the models presented in this 
work.
We simulate one quadrant of each model, using the technique described
and tested by Ercolano et al. (2003b), whereby the positive x-y, y-z and x-z planes
intersecting the z-, x- and y- axes respectively at zero, act as
mirrors, reflecting the incoming photons back into the simulated
cube.  The 
full volume is finally described by 10$^6$ cells for the shell models
and by 125000 cells for the spherical models ionised by 240 
sources. The number of energy packets used varies during the course 
of our simulations, but typically 1-10 million packets are sufficient 
for our grids to reach convergence within 10-20 iterations. We 
experimented with higher resolution grids and a larger number of 
energy packets and found our results to be virtually unaffected.

\subsection{Validation of our models}
\label{sec:vm}

\begin{figure*}
\begin{center}
\begin{minipage}[t]{17cm}
\includegraphics[width=17cm]{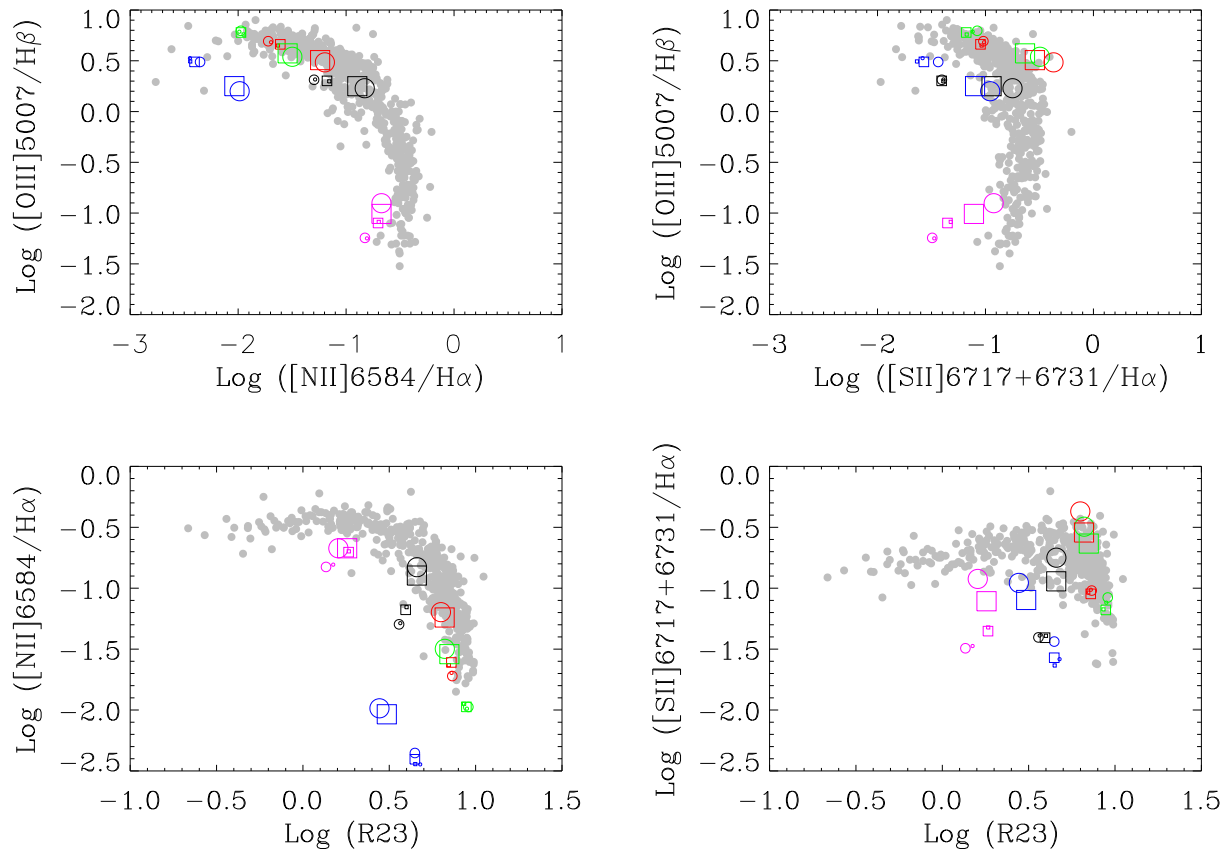}
\end{minipage}
\caption[]{H~{\sc ii} region excitation sequence in terms of $[$N~{\sc
     ii}$]$/H$\alpha$ and $[$S~{\sc ii}$]$/H$\alpha$ vs.$[$O~{\sc
     iii}$]$/H$\beta$ and R23.  The grey points represent giant 
H~{\sc ii} regions in galaxies   taken from  Garnett \& 
Kennicutt (1994),  Garnett et al. (1997), van Zee et al. (1998), 
Bresolin et al. (1999, 2004,  2005),  and Kennicutt et al. (2003), spanning
metallicities from 12~+~log(O/H)~$\sim$~8 to $\sim$~8.8., and
metal-poor emission line galaxies from Izotov et al. (2006), which
extend to 12~+~log(O/H)~$\sim$~7.2. The symbols and colors correspond 
to our models as defined in Table~\ref{tab:mods}. Please refer to the on-line version of this paper for colour figures.}
\label{fig:val1}
\end{center}
\end{figure*}

\begin{figure*}
\begin{center}
\begin{minipage}[t]{17cm}
\includegraphics[width=17cm]{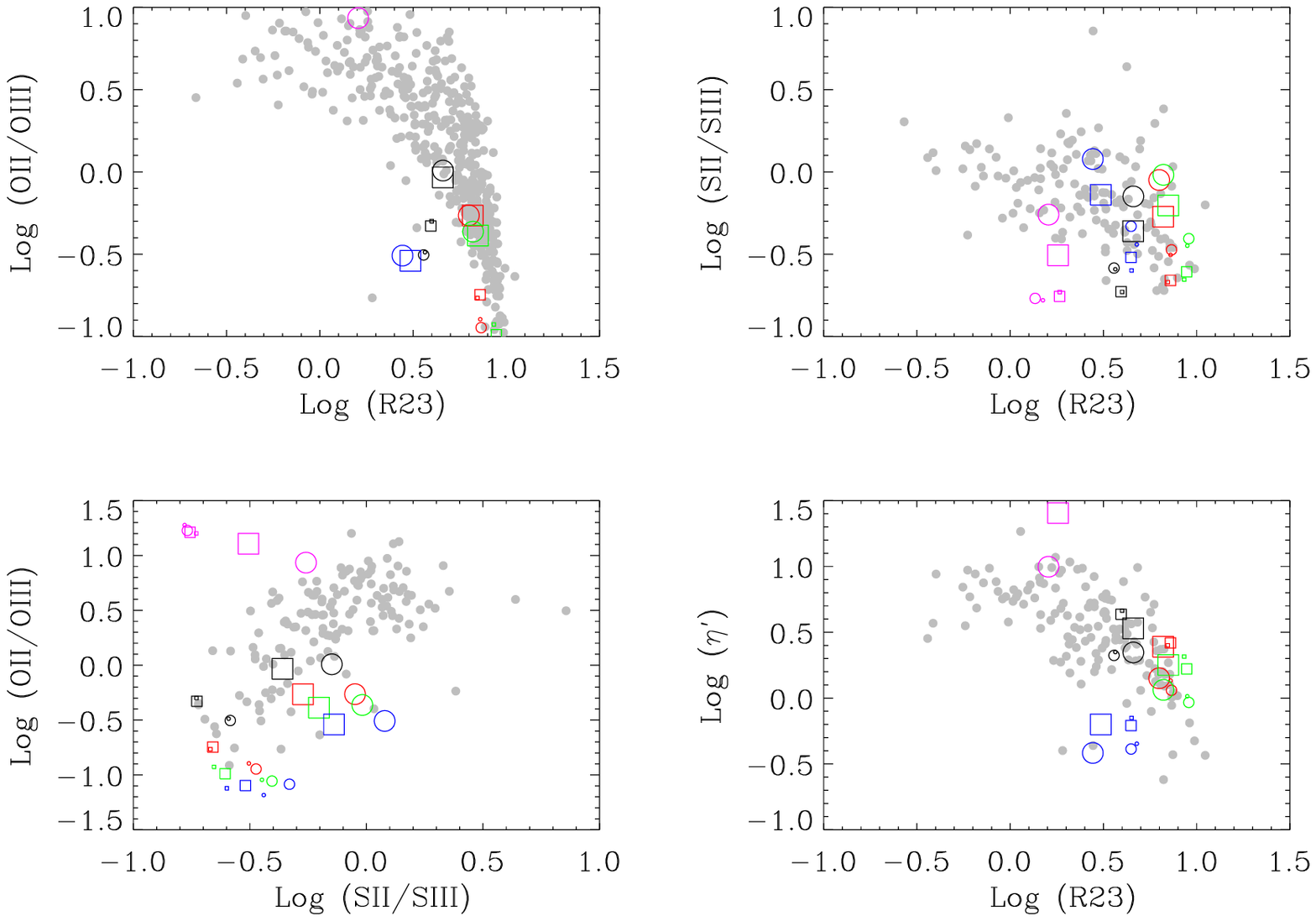}
\end{minipage}
\caption[]{H~{\sc ii} region ionisation sequence parametrised in 
terms of oxygen and sulphur. The grey dots indicate the position 
of giant H~{\sc ii} regions in spiral galaxies taken from  Garnett \& 
Kennicutt (1994),  Garnett et al. (1997), van Zee et al. (1998), 
Bresolin et al. (1999, 2004,  2005), Kennicutt et al. (2003), spanning
metallicities from 12~+~log(O/H)~$\sim$~8 to $\sim$~8.8. The top right
panel also includes a few points from metal-poor emission line galaxies from Izotov et al. (2006), which
extend to 12~+~log(O/H)~$\sim$~7.2; it was not possible to use this
data for the other three panels, due to the lack of simultaneous
detection of all the lines needed. 
 The symbols and colors correspond to our models as defined in
 Table~\ref{tab:mods}. Please refer to the on-line version of this paper for colour figures.
}
\label{fig:val2}
\end{center}
\end{figure*}

As stated above, the aim of this paper is not to provide new
calibrations to abundance diagnostic ratios nor to assess their
absolute accuracy. It is therefore not in our intention to create models
that fit any particular observations. While trying to maximise the
effects of a complex ionising field distribution, it is however still necessary to
ensure that the ionisation and temperature structures and, hence, emission 
line ratios we
obtain from our models are in the range of those observed in nature. In
Figure~\ref{fig:val1} we plot our results in a number of line ratio diagrams,
including those following Veilleux \&
  Osterbrock (1987) and Osterbrock, Tran \& Veilleux (1992), 
that show the variation in $[$N~{\sc ii}$]$ and $[$S~{\sc ii}$]$
excitation parametrised in terms of $[$O~{\sc iii}$]/H\beta$ (top
panels) and those following \citet{MRS85} showing the variation in
terms of R23\footnote{R23 = ($[$O~{\sc iii}$]$\,5007,4959 + $[$O~{\sc 
ii}$]$\,3726,3729)/H$\beta$}. Similarly, in Figure~\ref{fig:val2} we
plot the H~{\sc ii} region ionisation sequence parametrised in terms
of oxygen and sulphur line ratios \footnote{$\eta$'~=~($[$O~{\sc ii}$]$\,3726,3729/$[$O~{\sc iii}$]$\,5007,4959)/($[$S~{\sc ii}$]$\,6716,6731/$[$S~{\sc iii}$]$\,9069,9532)}.
The grey dots represent giant  H~{\sc ii} regions in spiral galaxies, 
taken from  Garnett \& Kennicutt (1994),  Garnett et al. (1997), van 
Zee et al. (1998), Bresolin et al. (1999, 2004,  2005),  and 
Kennicutt et al. (2003), spanning
metallicities from 12~+~log(O/H)~$\sim$~8 to
$\sim$~8.8. Figure~\ref{fig:val1} and the top right panel of
figure~\ref{fig:val2} also include a few points from
metal-poor emission line galaxies from Izotov et al. (2006), which
extend to 12~+~log(O/H)~$\sim$~7.2; it was not possible to use this
data for the other three panels of figure~\ref{fig:val2}, due to the lack of simultaneous
detection of all the oxygen and sulphur lines needed.
In general, our models fall within or near
the locus of observed H~{\sc ii} regions, with the exception of our
lowest metallicity ($Z/Z_{\odot}~=~0.05$) and highest
metallicity ($Z/Z_{\odot}~=~2$) models, which is not
surprising given that the observational sets available
did not include many data points at such extreme values of $Z$;
furthermore we did not attempt any systematic variation of 
the ionisation parameter with $Z$, while observations  of giant
H~{\sc ii} regions argue that the ionisation parameter decreases as $Z$ 
increases (see e.g. Mc Gaugh 1994). However this does not affect the
achievement of our aims.

\section{Results}
The temperature structure and emission line spectra obtained by our models
are analysed here in detail. Temperature fluctuations, which may 
introduce errors in the empirical
calculations of abundances, are examined.
The robustness of commonly used abundance diagnostics and ionic 
temperature relations
are also tested against possible errors introduced by geometrical effects.

\subsection{Temperature structure}
\label{sec:tst}

\begin{figure*}
\begin{center}
\begin{minipage}[t]{8.5cm}
\includegraphics[width=8.5cm]{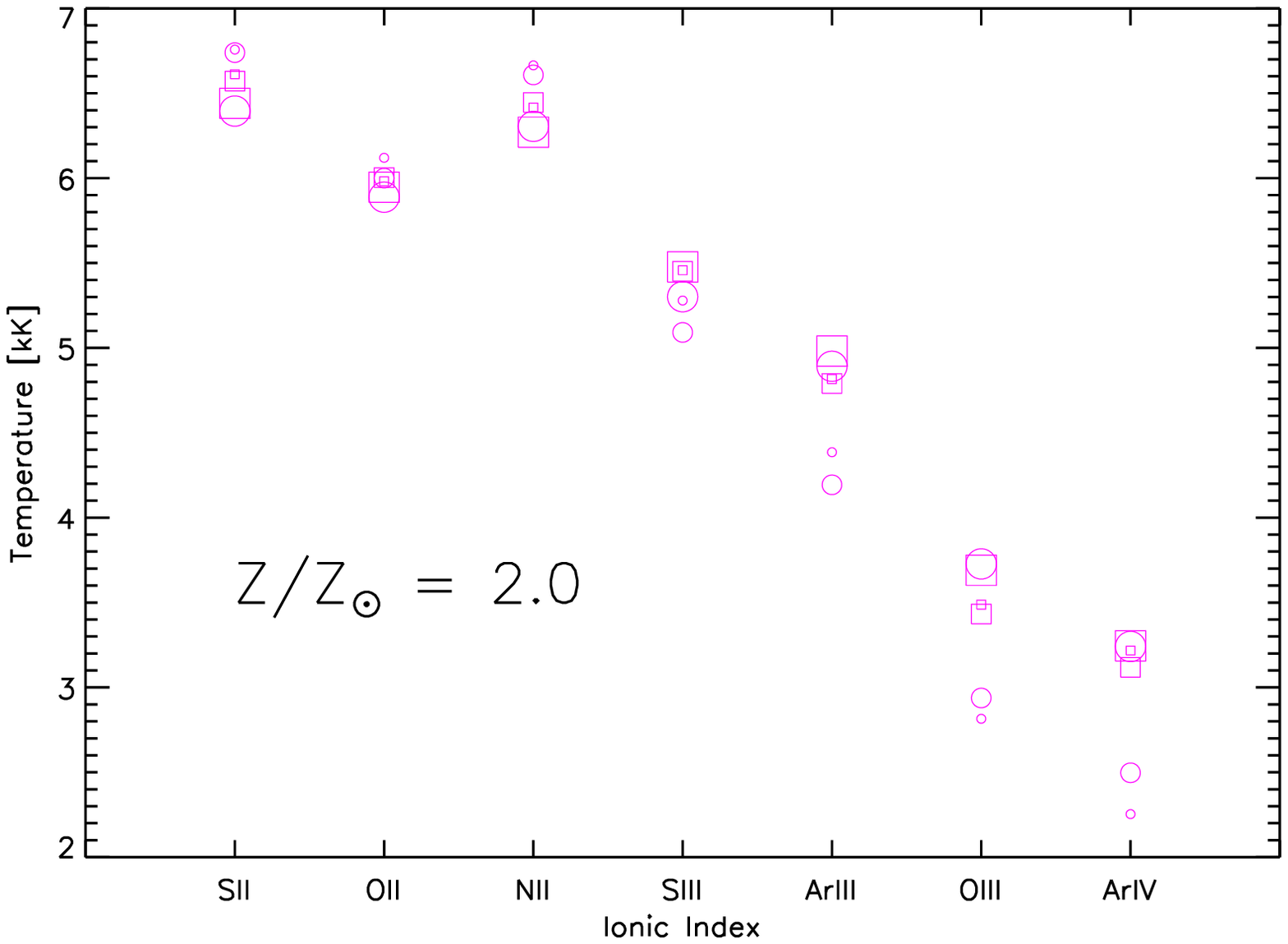}
\end{minipage}
\begin{minipage}[t]{8.5cm}
\includegraphics[width=8.5cm]{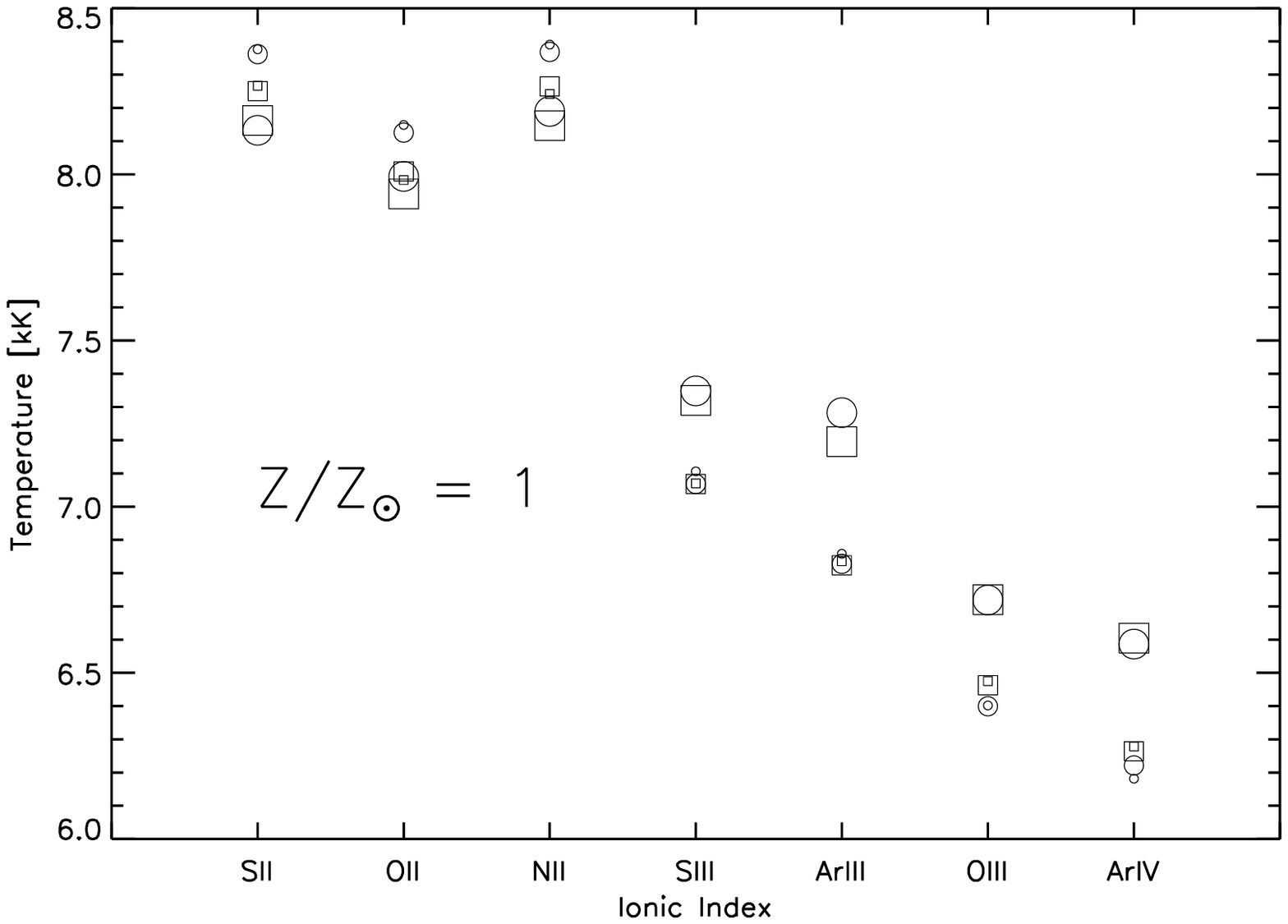}
\end{minipage}
\begin{minipage}[t]{8.5cm}
\includegraphics[width=8.5cm]{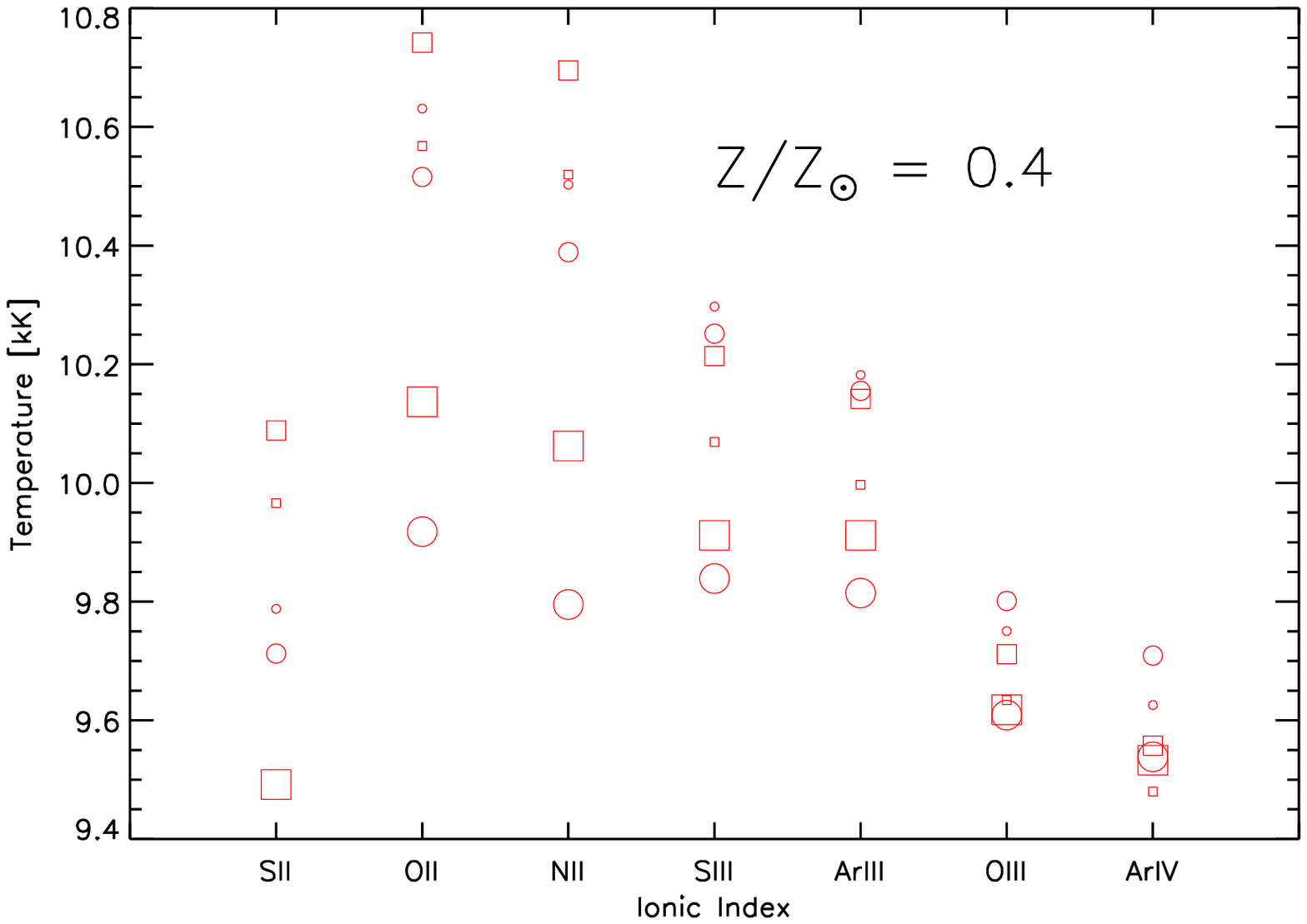}
\end{minipage}
\begin{minipage}[t]{8.5cm}
\includegraphics[width=8.5cm]{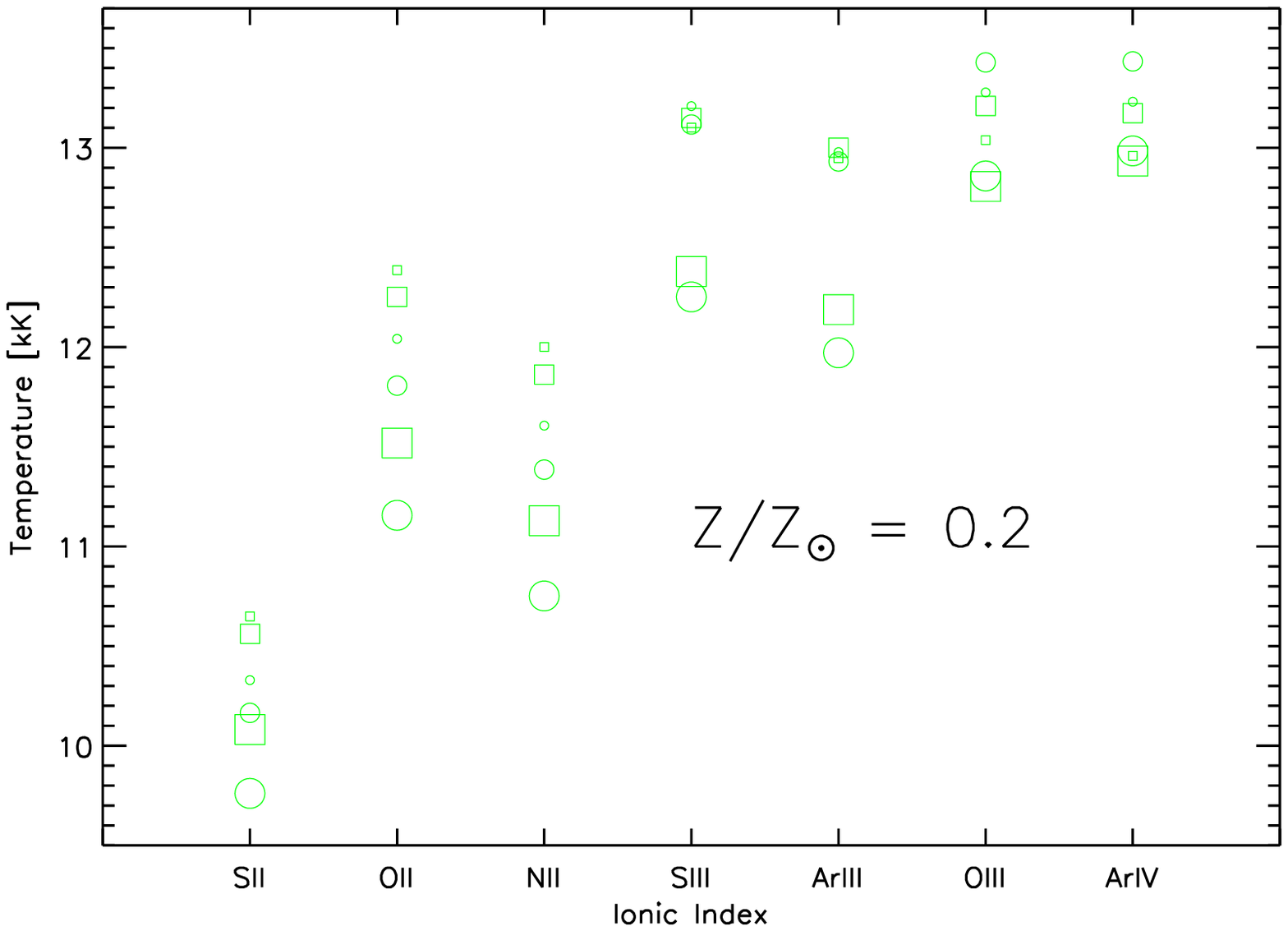}
\end{minipage}
\begin{minipage}[t]{8.5cm}
\includegraphics[width=8.5cm]{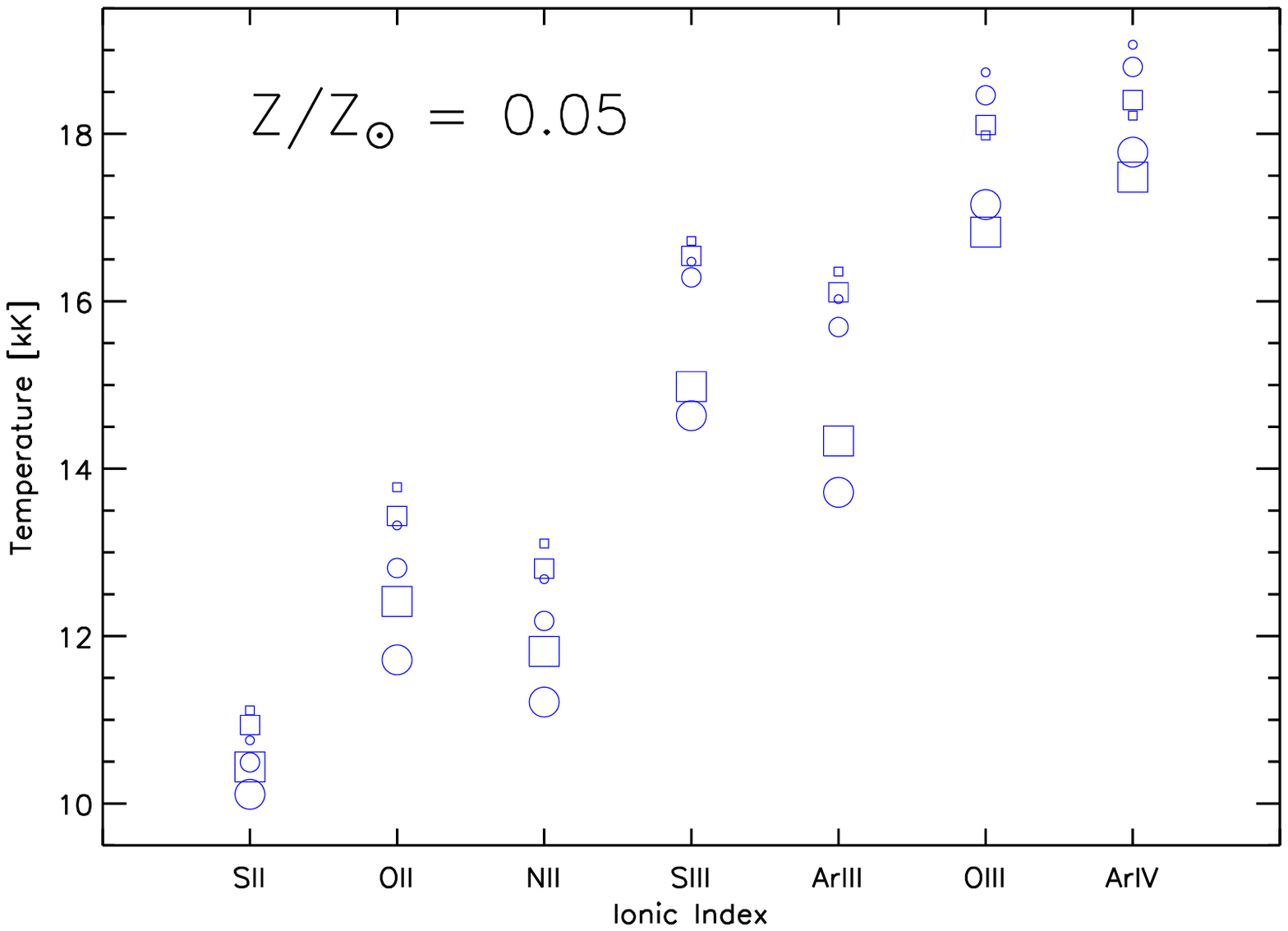}
\end{minipage}
\begin{minipage}[t]{8.5cm}
\includegraphics[width=8.5cm]{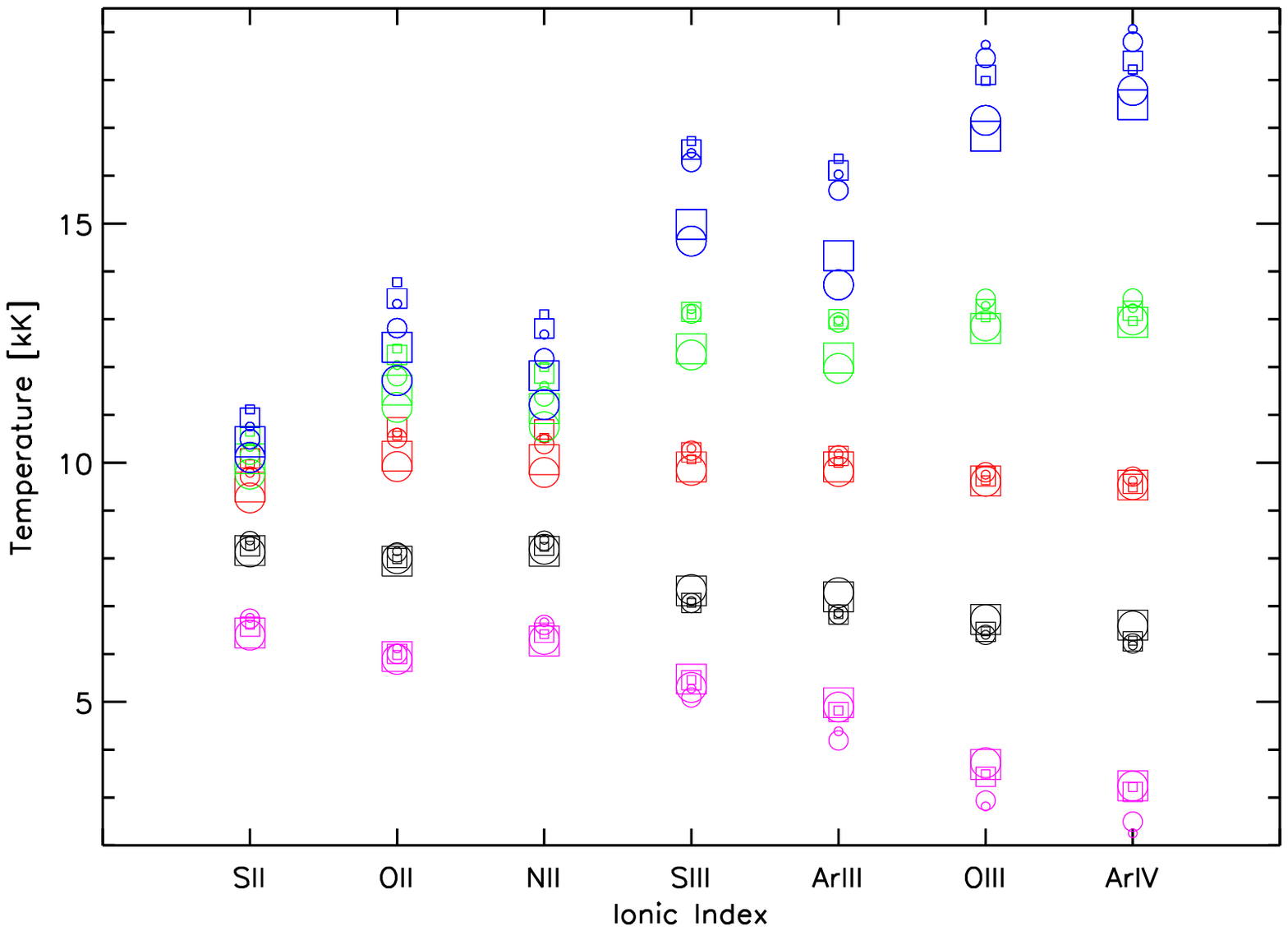}
\end{minipage}
\caption[]{Volume integrated electron temperatures weighted by ionic 
species. {\it Top left:} results for all models with 
$Z/Z_{\odot}$~=~2 (magenta); {\it top right:} results for all 
models at $Z/Z_{\odot}$~=~1 (black); {\it middle left:} results for 
all models at $Z/Z_{\odot}$~=~0.4 (red); {\it middle right:} 
results for all models at $Z/Z_{\odot}$~=~0.2 (green); {\it bottom 
left:} results for all models at $Z/Z_{\odot}$~=~0.05 
(blue); {\it bottom right:} Results for all models at 
all metallicities. A legend for the symbols is given in 
Table~\ref{tab:mods}. Please refer to the on-line version of this paper for colour figures.}
\label{fig:temps}
\end{center}
\end{figure*}

\begin{table*}
\begin{center}
\caption{Mean temperatures and temperature fluctuations quantified in terms of $T_0$(H$^+$), $t^2$(H$^+$),
 $T_0$(O$^{+}$), $t^2$(O$^{+}$),$T_0$(O$^{2+}$) and $t^2$(O$^{2+}$) (Peimbert, 1967)}
\begin{tabular}{lcccccc}
\hline										
model & $T_0(H^+)$ & $t^2(H^+)$ & $T_0(O^+)$ & $t^2(O^+)$ &$T_0(O^{+2})$ & $t^2(O^{+2})$ \\
\hline										
CSphH0.05   &   18240&    0.0178&   13350&    0.041&   18500&   0.007\\
HSphH0.05   &   17690&    0.0228&   12830&    0.046&   18500&   0.007\\
FSphH0.05   &   14850&    0.0511&   11720&    0.035&   17180&   0.009\\
            &       &$\pm$ 0.0006&     & $\pm$ 0.003&     &$\pm$ 0.004 \\
CSphH0.2    &   13140&    0.0034&   12060&    0.0145&   13430&   0.0010\\
HSphH0.2    &   13230&    0.0036&   11810&    0.0156&   13430&   0.0005\\
FSphH0.2    &   12100&    0.0112&   11160&    0.0141&   12860&   0.0010\\
              &       &$\pm$ 0.0001&     & $\pm$ 0.0004&     &$\pm$ 0.0001\\
CSphH0.4   &     9850&   0.0063&   10620&   0.0069&    9790&   0.0053\\
HSphH0.4   &     9860&   0.0038&   10500&   0.0060&    9790&   0.0029\\
FSphH0.4   &     9710&   0.0036&    9910&   0.0047&    9600&   0.0024\\
              &      &$\pm$ 0.0005&    & $\pm$ 0.0007&     &$\pm$ 0.0003 \\
CSphH1.0    &    6690&    0.024&    8130&    0.020&    6380&   0.012\\
HSphH1.0     &    6670&    0.021&    8110&    0.020&    6380&   0.009\\
FSphH1.0     &    7320&    0.018&    7990&    0.012&    6710&   0.009\\
              &       &$\pm$ 0.006&     & $\pm$ 0.002&     &$\pm$ 0.001 \\
CSphH2.0     &    5000&     0.156&    6110&    0.043&    2910&   0.130\\
HSphH2.0     &    4880&     0.152&    5980&    0.050&    2919&   0.096\\
FSphH2.0     &    5360&     0.075&    5890&    0.036&    3700&   0.084\\
              &        &$\pm$ 0.005&     & $\pm$ 0.003&     &$\pm$ 0.003\\
CShhH0.05 &   17510&    0.0129&   13770&    0.0437&   18110&   0.0041\\
HShhH0.05 &   17520&    0.0166&   13440&    0.0438&   18110&   0.0057\\
FShhH0.05 &   15200&    0.0369&   12410&    0.0348&   16840&   0.0098\\
              &       &$\pm$ 0.0002&     & $\pm$ 0.0002&     &$\pm$ 0.0001 \\
CShhH0.2   &   12940&   0.0033&   12390&    0.0113&   13210&   0.0017\\
HShhH0.2   &   13080&   0.0029&   12250&    0.0124&   13210&   0.0010\\
FShhH0.2   &   12280&   0.0082&   11520&    0.0121&   12810&   0.0015\\
              &      &$\pm$ 0.0002&     & $\pm$ 0.0002&     &$\pm$ 0.0007 \\
CShhH0.4   &     9780&   0.0091&   10570&   0.0089&    9710&   0.0078\\
HShhH0.4   &     9880&   0.0084&   10740&   0.0081&    9710&   0.0068\\
FShhH0.4   &     9810&   0.0060&   10130&   0.0064&    9620&   0.0046\\
              &       &$\pm$ 0.0003&    & $\pm$ 0.0003&     &$\pm$ 0.0001 \\
CShhH1.0    &    6880&    0.030&    7980&    0.026&    6460&    0.017\\
HShhH1.0    &    6850&    0.029&    8010&    0.026&    6460&    0.016\\
FShhH1.0    &    7280&    0.021&    7940&    0.017&    6720&    0.012\\
              &       &$\pm$ 0.001&     & $\pm$ 0.001&      &$\pm$ 0.001 \\
CShhH2.0    &    5450&    0.086&    5980&    0.045&    3420&   0.09\\
HShhH2.0    &    5430&    0.090&    6000&    0.045&    3420&   0.09\\
FShhH2.0    &    5490&    0.070&    5950&    0.037&    3690&   0.08\\  
              &       &$\pm$ 0.006&     & $\pm$ 0.006&     &$\pm$ 0.01 \\
\hline
\end{tabular}
\end{center}
\label{tab:t2}
\end{table*}

The volume integrated electron temperatures weighted by a number of 
commonly observed ionic species
are plotted in Figure~\ref{fig:temps} for metallicities of 
$Z/Z_{\odot}$~=~0.05, 0.2, 0.4, 1 and 2. The values were obtained 
according to
\begin{equation}
T_e({\rm X^{+i}}) = \frac{\int N_{\rm e} N({\rm X^{+i}}) T_{\rm e} \, 
dV}{\int N_{\rm e} N({\rm X^{+i}}) \, dV}
\label{eq:meantempferl}
\end{equation}
The models are represented by circles and squares of various 
colours and sizes as described in Table~\ref{tab:mods}.

  At a first glance some trends are apparent. At 
solar and high metallicities (i.e. $Z/Z_{\odot}~\ge~$1) the electron 
temperatures in 'high' ionic species zones (e.g. O~{\sc iii}, S~{\sc 
iii}, Ar~{\sc  iii} and N~{\sc iii}) are higher for models in cases H 
and F. The opposite is true for electron temperatures in 'low' ionic 
species zones, such as
O~{\sc ii} and  N~{\sc ii}. This implies a shallower temperature
gradient for case F and H when compared to case C, since at these
metallicities the highest nebular temperatures are reached in the low
ionisation zones which are not affected by the very efficient cooling
provided by the IR fine-structure lines of $[$O~{\sc iii}$]$ as
in the high ionisation zones, where, in fact, the lowest temperatures
are reached. The $Z/Z_{\odot}$~=~0.05 and 0.2 models do not follow the same
trend; here there is a monotonic shift of temperatures, whereby C
models are hotter than the H and F models, regardless of ionic species
zone. At intermediate metallicities ($Z/Z_{\odot}$~=~0.4) we are in a 
transition case from the two separate behaviours described above.
At low metallicities, the cooling is dominated by collisional 
excitation of H Ly$\alpha$, which increases as one gets closer to the 
ionisation front, where the proportion of residual neutral hydrogen 
increases. This results in a  outward decreasing temperature, 
contrary to the high metallicity cases. 

Differences in the temperature structure of cases F and C can be 
understood as follows: at each point in the nebula the electron 
temperature is primarily determined by the average
energy of the photons absorbed by H and He and by the cooling
efficiency of the ions. The latter is naturally less important at
lower metallicities. However, for the intermediate to high metallicity
cases, it is differences in the distribution of the cooling ions 
rather than differences in the heating rates which explains the 
difference in temperature structure between case C and case H for
higher and intermediate metallicities.

\subsection{Temperature Fluctuations}
\label{sec:t2}

\begin{figure*}
\begin{center}
\begin{minipage}[t]{17.cm}
\includegraphics[width=17cm]{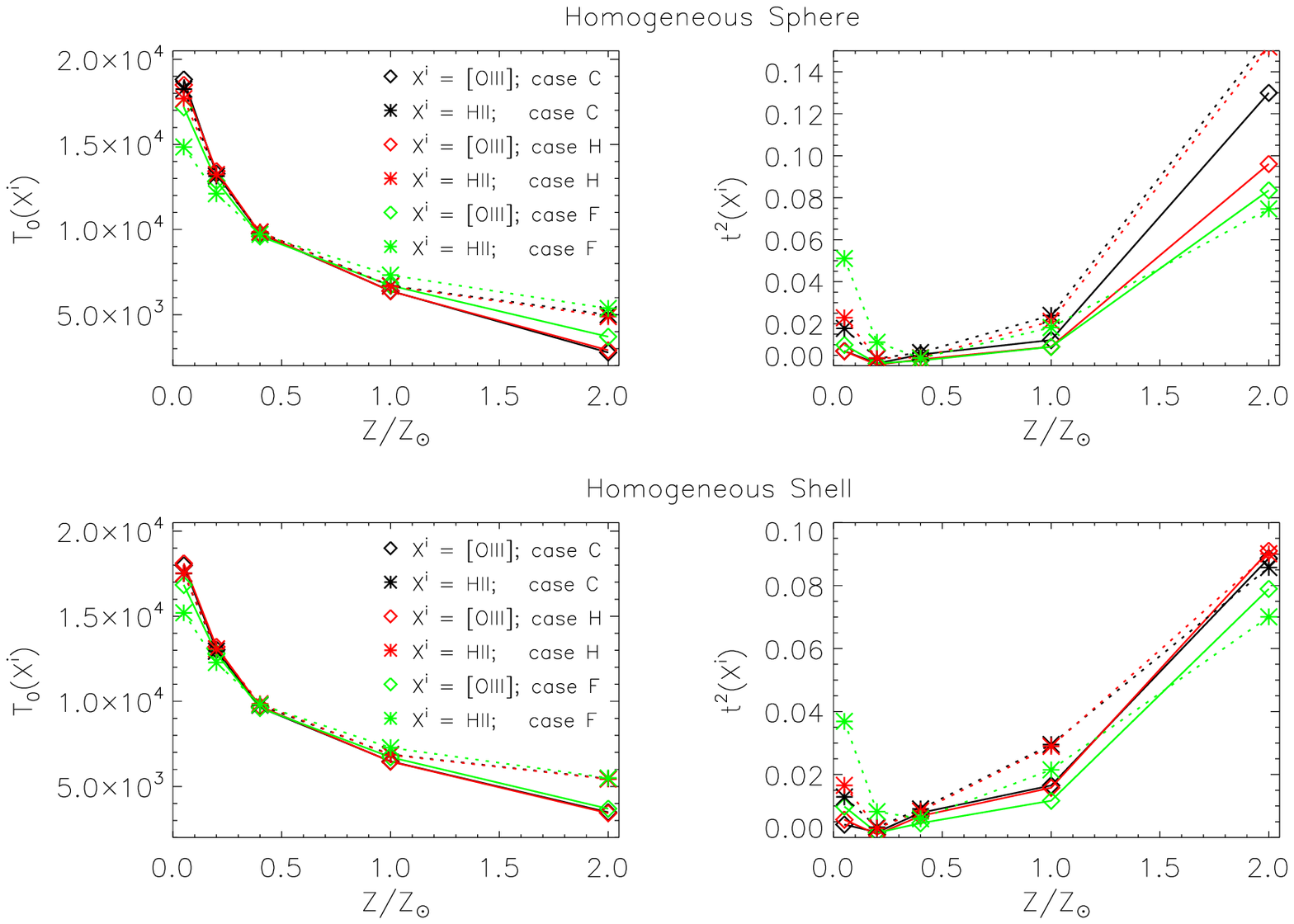}
\end{minipage}
\caption[]{ Mean ionic temperatures and temperature fluctuations as a function of gas
   metallicity for the homogeneous sphere and shell models. 
$t^2(O^{2+})$ and $t^2(H~^{+})$ are
   represented by solid and dashed lines, respectively. The lines are
   colour coded such that black, red and green are for cases C, H and
   F, respectively. Please refer to the on-line version of this paper for colour figures.}
\label{fig:t2hom}
\end{center}
\end{figure*}


We have calculated for the main ionic species in our models the formal values of the mean ionic temperatures, T$_0$, and of the 
temperature fluctuations,  $t^2$, according to the formalism of
Peimbert (1967).  Note that in the case of 
concentric sources, $t^2$ measures changes in radial  gradients of 
the temperature rather than ``temperature fluctuations'' while in the 
cases of distributed sources it really represent a temperature 
fluctuation.  For brevity we only list the 
values for H$^+$, O$^+$ and O$^{2+}$ in Table~\ref{tab:t2}. 
The errors quoted in the table are representative of the
accuracy achieved by our models. They contain contributions from
the variance intrinsic to our Monte Carlo approach and the error
introduced by using a finite grid to describe the ionised
region. The errors were estimated by considering that for a fully
spherically symmetric case, such as one with homogeneous gas
density distribution and a central location for all ionising sources, the
$t^2$ in an infinitely narrow spherical shell centred on the source of
the ionising photons should be zero. We note that our errors
increase with increasing metallicities, this is due to the larger
temperature gradients occurring at higher metallicities causing the
error contribution due to the finite grid description to
increase. Clearly all errors could be reduced by increasing the
number of energy packets and grid cells used in the
simulations. However we note that our errors are always at least one
order of magnitude smaller than the $t^2$ values and therefore do not
affect our conclusions. Given the large number of 3D models run for
this work (which vastly exceeds those finally presented here), we feel that a good balance between accuracy and
computational expense was achieved.

The value of $t^2$(O$^{+}$) is very hard to derive empirically. We know however of three H~{\sc ii} regions where this measurement has been made: the Orion
nebula (Esteban et al. 2004), M20 (Garc\'ia-Rojas et al., 2006) and M8
(Garc\'ia-Rojas et al. 2007). In those cases where $t^2$(O$^{+}$) cannot be derived empirically, it is 
generally assumed that $t^2$(O$^{+}$) = $t^2$(O$^{2+}$); it is clear
from the values listed in our table that this assumption is very often
 not verified and care should be taken to account for this in the error
estimation from such studies. For our $Z/Z_{\odot}$~=~2 models,
$t^2$(O$^{2+}$) is always a factor of 2 or more higher than
$t^2$(O$^{+}$), while for lower metallicities $t^2$(O$^{2+}$) becomes
lower than $t^2$(O$^{+}$) sometimes by large factors (up to
approximately 10).
  Finally, we  note that the formal $t^2$ values may diverge from
the empirical ones (e.g. Kingdon \& Ferland, 1995; Zhang, Ercolano \& Liu,
2007), that are generally based on the comparison of the
electron temperature derived from the depth of the Balmer jump and the
$[$O~{\sc iii}$]$ temperature, the values listed in
Table~\ref{tab:t2} are, nevertheless, sufficient to identify the effects of the
stellar distribution on temperature fluctuations in the nebular gas.

 In Figure~\ref{fig:t2hom} we plot the $t^2$ values
for H$^+$ and O$^{2+}$ (dashed and solid lines, respectively) against
metallicity for the spherical and shell models. Cases C, H and F
are represented by black, red and green lines and symbols
respectively. Unsurprisingly, the temperature fluctuations, which in
this case are a direct consequence of large-scale temperature
gradients, are larger for the high metallicity models. A full discussion of
this effect and of the causes of the large scale temperature fluctuations
in metal rich nebulae is given by Stasi\'nska (1980) and Kingdon \& 
Ferland (1995). What had not been noticed before, however, is that at 
very low metallicities, the $t^2$ values for  H$^+$ and O$^{+}$ rise
again; this is due to the large temperature gradient already shown in
Figure~\ref{fig:temps}. However, $t^2$(O$^{2+}$) remains tiny, so that the overall 
effect on abundance determinations is expected to be small.  

With regards to a comparison of the $t^2$ values obtained with the three different
spatial distributions of sources (cases C, H and F), we first of all
notice that for intermediate to high metallicities ($Z/Z_{\odot}~\geq~0.4$), 
case F models show a more isothermal gas (smaller $t^2$) than case H
or C models. This is due to the fact that the contributions to the
$t^2$ due to the true temperature {\it fluctuations}
created by the two different stellar populations in case F (and in a
smaller degree in case H) are completely washed out by the large-scale
temperature {\it gradients} caused by the metals' cooling. This effect vastly
dominates at these metallicities, resulting in larger values of $t^2$
being obtained by case C models, which, as discussed in the previous
section (see Figure~4), have steeper temperature
gradients than their cases H and F counterparts.

The above is further confirmed by the fact for very low metallicities (Z/Z$_{\odot}~\leq~0.2$), case F models (green lines) show larger fluctuations than
case C and H models. This is because at these low metallicities, case
C, H and F all show similar large-scale temperature gradients (see
Figure~4); the values of $t^2$ are thus larger for case F models where
true temperature fluctuations are at play. However, once again we
point out that no large effects are expected on abundance
determinations due to the fact that $t^2$(O$^{2+}$) is small in all
cases. 


\subsection{Ionic temperature relations}
\label{sec:itr}

\begin{figure*}
\begin{center}
\begin{minipage}[t]{8.5cm}
\includegraphics[width=8.5cm]{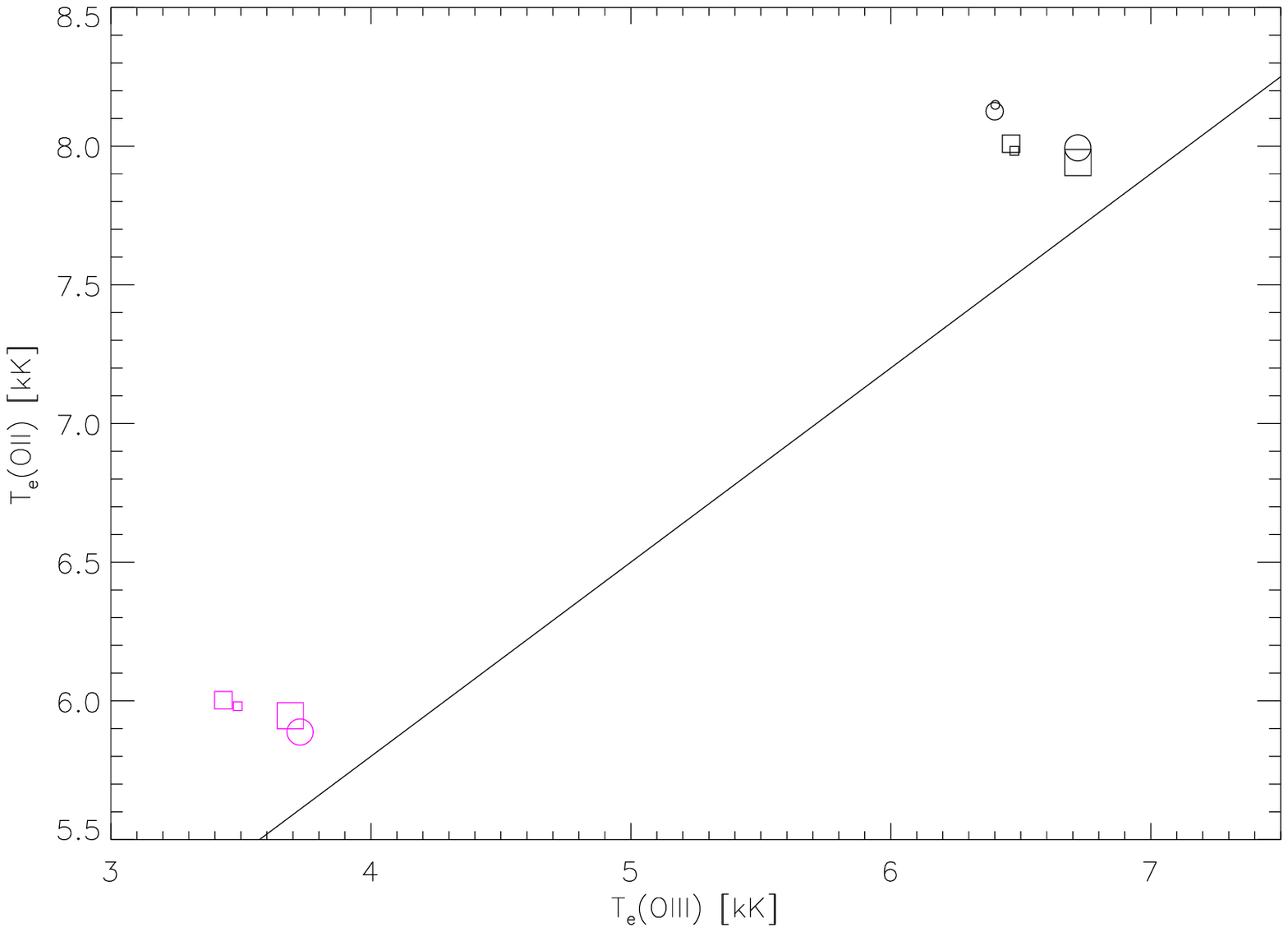}
\end{minipage}
\begin{minipage}[t]{8.5cm}
\includegraphics[width=8.5cm]{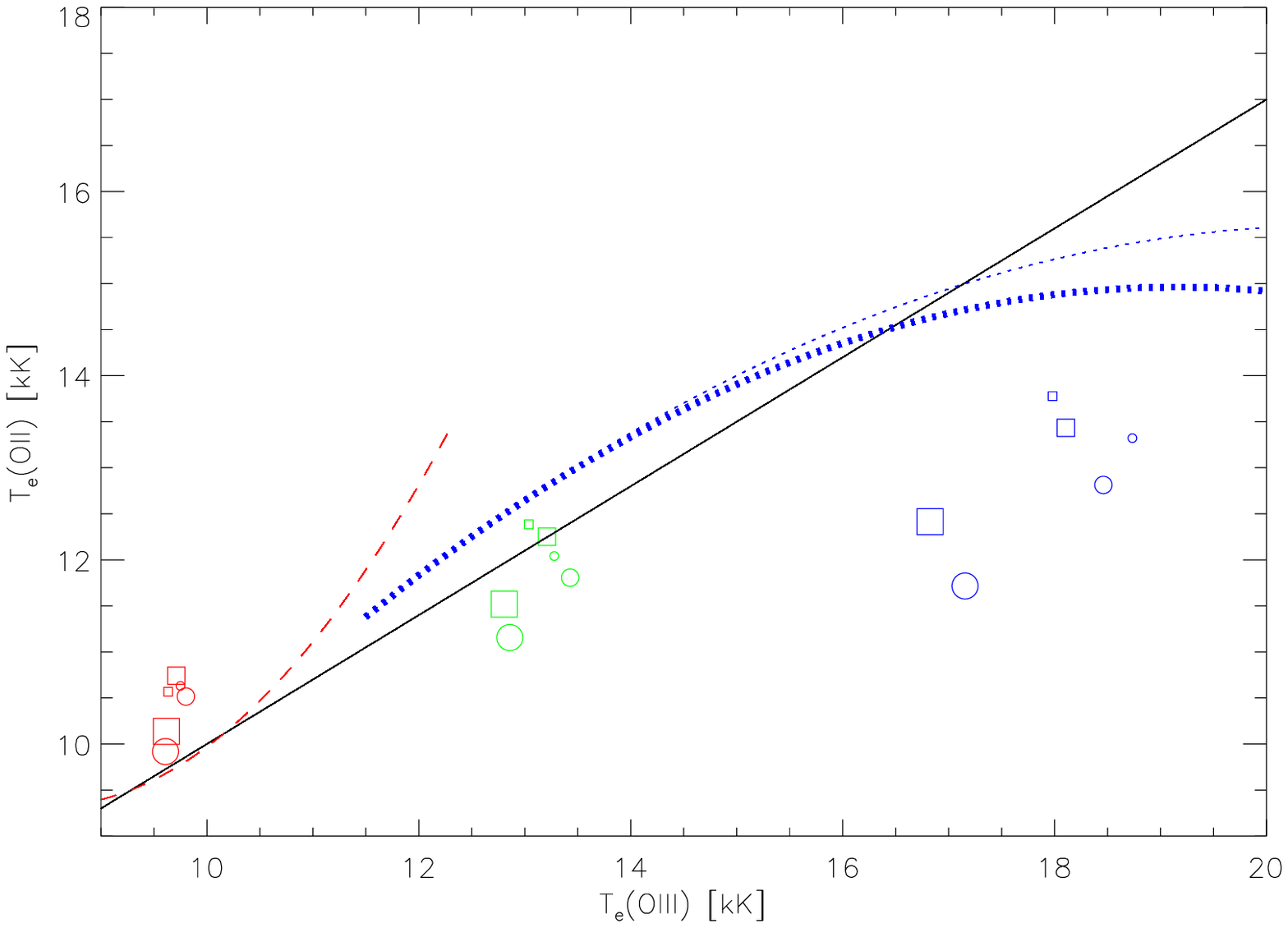}
\end{minipage}
\caption[]{T$_e$(O~{\sc ii}) versus
T$_e$(O~{\sc iii}) for Z/Z$_{\odot}$~=1.0 and 2.0 {\it left} and 
Z/Z$_{\odot}$~=0.05, 0.2 and 0.4 {\it right} models
for cases C, H and F. The straight black solid line is the scaling law of Garnett (1992)
and the blue dotted and red dashed curves are the relations given in Eqn~14 of
Izotov et al. (2006) for 12~+~log(O/H)~=~8.2 (red), 7.6 (blue dotted, thick)
and 7.2 (blue dotted, thin). The symbols representing our models are described in Table~\ref{tab:mods}. Please refer to the on-line version of this paper for colour figures.}
\label{fig:temprel}
\end{center}
\end{figure*}

\begin{table*}
\begin{center}
\caption{Oxygen abundances derived applying various metallicity 
indicators to the integrated emission line spectra
calculated for our models (See text for calibrations used). All abundances are given in logarithmic 
scale, 12~+~log(O/H). The input values of 12~+~log(O/H) (i.e. 'the right
answer') are 8.99 (Z/Z$_{\odot}$~=~2), 8.69 (Z/Z$_{\odot}$~=~1), 8.29
(Z/Z$_{\odot}$~=~0.4), 7.99 (Z/Z$_{\odot}$~=~0.2) and 7.39
(Z/Z$_{\odot}$~=~0.05). $\mid\Delta\mid$, the largest difference between the abundance derived from C,
H and F cases are given for each model trio. The value averaged over 
all models, $\mid<\Delta>\mid$, is given in the last row of the table.}
\begin{tabular}{lcccccclcccccc}
\hline
   model & O$_{23}$ &  O$_3$N$_2$   & N$_2$  &  S$_{23}$ & S$_3$O$_3$ 
& Ar$_3$O$_3$ &  model &  R$_{23}$ &  O$_3$N$_2$   & N$_2$  & 
S$_{23}$ & S$_3$O$_3$    & Ar$_3$O$_3$ \\
\hline										
CSp2.0 &  8.67&    8.78 &   8.44  &  7.92 & 8.77 &   8.83   &CSh2.0 & 8.62&    8.77&    8.50 &   8.09&    8.75&    8.82 \\
HSp2.0 &  8.69&    8.78&    8.43 &   7.87&  8.76&    8.81   &HSh2.0 & 8.62&    8.77&    8.50 &   8.08&    8.75&    8.82 \\
FSp2.0 &  8.67&    8.73&    8.52 &   8.17&  8.67&    8.78   &FSh2.0 & 8.63&    8.76&    8.52 &   8.16&    8.72&    8.81 \\
$\mid\Delta\mid$  & 0.02 &  0.05 &  0.09 &  0.30 &  0.10 &  0.05  &$\mid\Delta\mid$ &  0.01 &  0.01 &  0.02 &  0.08 &  0.03 &  0.01\\
\noalign{\smallskip}
CSp1.0 &8.74&    8.27&    8.17 &   7.76&    8.04&    8.48   &CSh1.0 &8.66&    8.30&    8.24 &   7.94&    8.16&    8.55  \\
HSp1.0 &8.75&    8.27&    8.16 &   7.73&    8.02&    8.46   &HSh1.0 &8.67&    8.30&    8.23 &   7.92&    8.14&    8.54  \\
FSp1.0 &8.49&    8.40&    8.43 &   8.34&    8.24&    8.62   &FSh1.0 &8.51&    8.38&    8.39 &   8.23&    8.24&    8.61   \\
$\mid\Delta\mid$ & 0.26 &  0.13 &  0.27 &  0.61 &  0.22 &  0.16 &$\mid\Delta\mid$ &  0.16 &  0.08 &  0.16 &  0.31 &  0.10 &  0.07 \\
\noalign{\smallskip}
CSp0.4 &8.52&    8.07&    7.87 &   7.93&    8.28&    7.96   &CSh0.4 &8.52&    8.09&    7.97 &   8.49&    8.10&    7.77  \\
HSp0.4 &8.52&    8.06&    7.85 &   7.92&    8.22&    7.91   &HSh0.4 &8.50&    8.10&    7.98 &   8.45&    8.08&    7.73  \\
FSp0.4 &8.43&    8.24&    8.15 &   8.22&    9.02&    8.25   &FSh0.4 &8.41&    8.23&    8.19 &   8.92&    8.27&    8.12  \\
$\mid\Delta\mid$ &   0.09 &  0.18 &  0.30 &  0.30 &  0.80 &  0.34 &$\mid\Delta\mid$ & 0.11 &  0.14 &  0.22 &  0.47 &  0.19 &  0.39 \\
\noalign{\smallskip}
CSp0.2 & 7.86&    7.97&    7.77 &   7.99&    7.67&    7.67   &CSh0.2 &7.84&    7.99&    7.79 &   8.18 &  7.87 &    7.92  \\
HSp0.2 & 7.87&    7.97&    7.77 &   8.01&    7.66&    7.66   &HSh0.2 &7.86&    7.97&    7.77 &   8.10 &  7.79 &    7.82  \\
FSp0.2 & 7.68&    8.15&    8.05 &   8.68&    8.10&    8.26   &FSh0.2 &7.72&    8.13&    8.02 &   8.61 &  8.11 &    8.26  \\
$\mid\Delta\mid$ &  0.19 & 0.18 &  0.28 &  0.69 &  0.44 &  0.60 &$\mid\Delta\mid$ &  0.14 & 0.16 &  0.25 &  0.51 &  0.32 &  0.44\\
\noalign{\smallskip}
CSp0.05&7.48&    7.92&    7.51 &   7.29&    7.39&    6.30   &CSh0.05& 7.44&    7.93&    7.51 &   7.38&    7.58&    6.67  \\
HSp0.05&7.44&    7.95&    7.56 &   7.37&    7.49&    6.51   &HSh0.05& 7.44&    7.94&    7.53 &   7.38&    7.57&    6.63  \\
FSp0.05&7.15&    8.12&    7.77 &   7.73&    7.91&    7.48   &FSh0.05& 7.21&    8.09&    7.74 &   7.69&    7.93&    7.49  \\
$\mid\Delta\mid$ &  0.33 &  0.20 &  0.26 &  0.44 &  0.52 &  1.18 &$\mid\Delta\mid$ & 0.23 &  0.16 &  0.23 &  0.31 &  0.36 &  0.86\\
\noalign{\smallskip}
\hline
  $<\mid\Delta\mid>$ &0.18 & 0.15&  0.24& 0.47 & 0.41 & 0.47 & $<\mid\Delta\mid>$ &0.12 & 0.11& 0.18 & 0.32& 0.23 & 0.45\\
\hline
\end{tabular}
\label{tab:abun}
\end{center}
\end{table*}

We have shown that the temperature structure of models with centrally 
concentrated ionising sources, case C, may vary compared to those of similar models 
where the sources are distributed within the half volume (case H),
which have partially overlapping Str\"omgren spheres, and
to those with sources distributed within the full volume (case F),
which have fully independent Str\"omgren spheres. Case C to case F variations in the temperature
structure of the models may have implications for a number of 
ionic temperature relations. These scaling laws are often employed in abundance
studies when observational data for a given ionic zone is missing. Some of the most popular relations we found in the
literature include T$_e$(S~{\sc iii}) vs
T$_e$(Ar~{\sc iii}), T$_e$(O~{\sc iii}) vs T$_e$(Ar~{\sc iii}),  
T$_e$(O~{\sc ii}) vs T$_e$(N~{\sc ii}). These
ratios are expected to be around unity and we found them to be very little affected by the shifts in
temperatures due to the spatial distribution of stars. This is not
surprising, given that the ionic species involved are both 'high' (O~{\sc iii}, S~{\sc 
iii}, Ar~{\sc  iii} and N~{\sc iii})  or 'low' (O~{\sc ii} and  N~{\sc
  ii}), and therefore while the absolute
temperature values in each case may shift to higher or lower values,
the resulting ratios remain virtually unaffected.

Scaling laws have also been derived for 
T$_e$(O~{\sc iii}) vs 
T$_e$(S~{\sc iii}) and
T$_e$(O~{\sc iii}) vs T$_e$(O~{\sc ii}), by Garnett (1992), based on his 
own  grids of photoionisation models and those by Stasi\'nska 
(1982). 
More recently, Izotov et al. (2006) produced somewhat different relations 
between T$_e$(O~{\sc ii}) and T$_e$(O~{\sc iii}), and between
T$_e$(S~{\sc iii}) 
and T$_e$(O~{\sc ii}) (their Eqs. 14 and 15, valid only for metal poor
cases to 12~+~log(O/H)~$\sim$~8.2), based on a set 
of up-to-date (but still spherically symmetric) photoionisation models 
that reproduce the observed trends of metal-poor galaxies. They found 
that the observed values of T$_e$(O~{\sc ii}), T$_e$(O~{\sc iii}) and T$_e$(S~{\sc iii}) in a large 
sample of H~{\sc ii} galaxies reproduce the theoretical relations, but with 
a large scatter (not only attributable to observational errors in the 
case of T$_e$(O~{\sc ii}), T$_e$(O~{\sc iii}), see their Figures~4a and b). 

Figure~\ref{fig:temprel} shows the
behaviour of our Z/Z$_{\odot}$~=1.0 and 2.0 (left panel, black and
magenta symbols respectively) and
Z/Z$_{\odot}$~=0.05, 0.2 and 0.4 (right panel, blue, green and red
symbols, respectively) metallicity models
for cases C, H and F with regards to the T$_e$(O~{\sc ii}) versus
T$_e$(O~{\sc iii}) scaling laws of Garnett (1992, black solid line in
both panels)
and Izotov et al. (2006, coloured curves in the right panels). The symbols are as described in Table~\ref{tab:mods}. Our lowest metallicity bin
Z/Z$_{\odot}$~=~0.05 (corresponding to 12~+~log(O/H)=~7.39) falls in between
Izotov's 12~+~log(O/H)=~7.2 and 7.6 metallicity bins. Their T(O~{\sc
  iii}) versus T(O~{\sc ii})
relations given in their Eqns~14 are therefore plotted in blue in
the right panel of our Figure~\ref{fig:temprel}, with the
thinner and thicker curves indicating the 7.2 and 7.6 metallicity bin,
respectively. 
Our Z/Z$_{\odot}$~=~0.4 metallicity bin (red symbols) corresponds to
12~+~log(O/H)=~8.29, slightly above Izotov's 12~+~log(O/H)~=~8.2 bin 
which, nevertheless is represented by the red line in Figure~\ref{fig:temprel}.

A 'mixed' relation such as T$_e$(O~{\sc iii}) vs T$_e$(O~{\sc ii})  is much more affected  by the fact that
 low and high ionic temperatures shift in opposite directions for
 intermediate (red points; Z/Z$_{\odot}$~=~0.4) and high metallicities
 (magenta and black points; Z/Z$_{\odot}$~=~2 and 1). This will
 contribute to the scatter noticed observationally for this relation. 
For example, Kennicutt, Bresolin \& Garnett (2003) presented a study of those H~{\sc ii} regions 
in M~101 for which direct measurements
of the nebular auroral lines could be obtained. They found the
T$_e$(O~{\sc iii}) vs T$_e$(S~{\sc iii}) relation
to be matched closely by their observations, whereas the T$_e$(O~{\sc
   iii}) and T$_e$(O~{\sc ii}) temperatures turned out to be 
rather uncorrelated. It is also worth noting at this point tht another factor that could
contribute to the scatter in the  T$_e$(O~{\sc iii}) vs T$_e$(O~{\sc ii})
relation is the
 dependency of T$_e$(O~{\sc ii}) on the electron density (P\'erez-Montero \& 
D\'{i}az
 2003). The scatter found by our models is for a given electron density,
and therefore
it may be even larger for a different sample covering a wider range of
densities. The ionising flux distribution is certainly a
 factor that contributes to weakening the correlation, however, as pointed out by Stasi\'nska (2005), at high metallicities, the
temperature derived from the $[$O~{\sc ii}$]$ line ratio may by strongly in
error due to the contribution of recombination from O$^{++}$.

\subsection{Abundance diagnostics}

\label{sec:ad}


We have assessed that the geometrical distribution of stars within an
H~{\sc ii} region plays a role in the temperature structure of the
gas, the magnitude of these effects will clearly be dependent
on the total mass of the stars and the concentration level. It is
important now to verify the robustness of commonly used metallicity
indicator against these temperature shifts. In
particular we are interested in identifying systematic trends with
stellar distribution rather than absolute errors. Table~\ref{tab:abun}
lists the {\it empirical} oxygen abundances obtained using 6 different
indicators. We used the following calibrations for the 6 indicators:
(1) O$_{23}$ -- Pilyugin (2000, 2001b), (2) O$_{3}$N$_{2}$ -- Stasi\'nska
(2006), (3) N$_{2}$ -- Pettini \& Pagel (2004), (4) S$_{23}$ --
P\'erez-Montero \& D\'iaz (2005), (5) S$_{3}$O$_{3}$ -- Stasi\'nska 
(2006), (6) Ar$_{3}$O$_{3}$ -- Stasi\'nska (2006).
It is worth reiterating at this point that metallicity
indicators should, and generally are, only used for statistical 
studies as the error on a single region may be very large; the models 
presented in this work do not attempt to cover the full parameter space occupied by
  H~{\sc ii} regions and therefore do not aim at
identifying the most accurate indicator {\it in absolute terms}. Here 
we are mainly interested
in studying a possible systematic error in the derived abundances
introduced by the 3D stellar distribution. For this reason for each
trio of models (C, H and F for a given density distribution and
metallicity) we compute  $\Delta$, the largest difference between 
the oxygen abundance in log units derived from cases C, H and F. 

The results are summarised in Table~\ref{tab:abun}, where the mean values $<\Delta>$ 
are also listed. For all metallicity indicators apart from O$_{23}$, there is a clear trend for higher metallicity being
derived from models ionised by fully distributed sources, case F, than
from case C and H models. The reverse is true for O$_{23}$, where
metallicities derived from case F models are smaller than those
derived from case C and H models. The $\Delta$ values at a given
metallicity vary from one indicator to the next, however they are rarely below $\sim$0.1~dex
(only for the highest metallicity case), with more representative
values around 0.3~dex for the spherical case, but often larger than
0.4 (and larger than 1.1~dex in one case). These deviations are
slightly smaller for the shell density distributions; this is obvious
as even for F cases most of the ionising radiation in these models
will be emitted from the central cavity, reducing the differences
between C and F cases. We note that in some cases the values given 
by the case H models are lower than those given by the respective 
case C models (rather than being equal or in the middle between C and F). 
We have analysed these deviations statistically and can confirm that the 
small differences are simply due to the variance of our Monte Carlo models 
and do not bear any physical significance. 

The reason for the systematic effect we see in the {\it empirical} abundance
determinations can be understood by re-examining
Figure~\ref{fig:val2}, where the ionisation sequence of our model is
parametrised in terms of oxygen and sulphur emission lines. The C
(small symbols), H (medium symbols) and F (large symbols) form a
sequence, with the case F models generally showing a lower ionisation
parameter than case C and H models. The differences in the abundances
derived from case C to F with the various indicators clearly reflect
the different dependence of each indicator on the ionisation
parameter. This can be simply shown for an idealised system. 
The ionisation parameter of a pure-hydrogen spherical volume of gas with
number density $N_H$ ionised by a source emitting $Q_{H^0}$
hydrogen-ionising photons per second is defined as $U$~=~$Q_{H^0}$/(4$\pi N_H
R_s^2$), where $R_{s}$ is the Str\"omgren radius. 
In such a system, from the ionisation balance equation
(e.g. Osterbrock, 1989, eqn 2.19) $R_{s}$ is directly proportional to $Q_{H^0}^{1/3}$
and  to $N_H^{-2/3}$, neglecting the temperature
dependence of the hydrogen recombination coefficient. 
For simplicity we compare the ionisation parameter of the system
above, $U^C$, ionised by only one source, to
that of a similar system (F) ionised by $N_s$ identical sources each
emitting $Q_{H^0}$/$N_s$ hydrogen-ionising photons per second.
With these assumptions it is easy to show that the ionisation
parameter, $U^F$, for system F, measured at the Str\"omgren radius of each individual
source is simply related to $U^C$ by
\begin{equation}
U^F = U^C{\cdot}N_s^{\frac{-1}{3}}
\end{equation}
and therefore always smaller than  $U^C$ for $N_s > 1$

The magnitude of the errors in the metallicity derived by the strong
line methods are significant. It is, however, true that the effects
reported here represent a worst case scenario, and our extreme
assumptions on stellar populations create a large dispersion
in the resulting ionisation parameters, which in some cases exceeds
the observed range, as seen in
Figures~\ref{fig:val1} and \ref{fig:val2}. 
Aside from the magnitude of the $\Delta$ values, however, a worrying aspect is
the fact that the discrepancies between cases C, H and F are
systematic. This can have an impact on galactic metallicity
gradients determined via strong line methods, if these are calibrated
via photoionisation models. In fact, if
compact clusters (close to case C) and loose associations (close to
case F) are randomly distributed throughout a given galaxy, then the
systematic errors would only cause a larger scatter in the observed
metallicities, but, given sufficient number statistics, they would not affect the 
measured metallicity gradient. However if the ratios of compact
clusters to loose associations is somewhat dependent on the
galactocentric distance, 
then the systematic errors due to stellar geometrical distributions 
may indeed introduce a bias on the measured galactic metallicity gradient, 
if the abundances are obtained from strong line methods calibrated 
on ab-initio models which do not reproduce the observed excitation of H~{\sc ii} regions.
For example, recent work by Rosolowsky 
et al. (2007), presenting high
resolution molecular gas maps of M33, showed a truncation in the mass
distribution of giant molecular clouds (GMCs) at a galactocentric
distance of 4~kpc. A recent study on the demographics of young
star-forming clusters in
M33 by Bastian et al. (2007) also shows the same cut-off at 4~kpc for the
clusters detected. We could interpret this as tentative
evidence of different star formation environments from the centre to
the edge of M33, however, we prefer to postpone this discussion until
more compelling observational evidence becomes available.

\section{Conclusions}

Following our theoretical investigation on the effects of the spatial 
configuration of ionisation sources on the temperature
structure of H~{\sc ii} regions, we summarise our conclusions as follows:

\begin{enumerate}
\item For intermediate to high metallicities
  (0.4$~\leq~Z/Z_{\odot}~\leq$~2), 
for a given gas density distribution, abundance and ionising spectral
shape 
and intensity, a model with a central concentration of stars (case C) will
result in higher ionic temperatures for high ionisation species (O$^{2+}$, 
S$^{2+}$ etc.), compared to the same model with stars fully distributed 
within the volume (case F). The opposite is true for 'low' ionisation
species (e.g. O$^+$, N$^+$). This results in a shallower gradients in
the electron temperature distribution across the ionic species zones. 
\item Low metallicity models (Z/Z$_{\odot}~\leq$~0.2) do not show the
  temperature inversion from low to high ionic species zones, rather a
  shift in the temperature is experienced by all ionic species zones,
  resulting in case F models being cooler than case C and H models. 
\item At intermediate to high metallicity ($Z/Z_{\odot}~\geq~0.4$), models with stars distributed within the full volume are more 
isothermal (show lower $t^2$ values) than the same models with a central concentration of 
stars. These temperature ``fluctuations'' obtained for case
C models are a simple consequence of a large temperature gradient. Multiple ionising sources of different temperatures at central 
or non-central locations do not produce significant temperature 
fluctuations in the ionised gas of models with $Z/Z_{\odot}~\geq~0.4$.
\item At low metallicities ($Z/Z_{\odot}~\leq~0.2$), models with stars
  distributed within the full volume (case F) show {\it larger} $t^2$ values
  than the same models with a central concentration of stars. Here we
  are seeing the effects of true temperature ``fluctuations'' for case
  F models. The magnitude of $t^2(O^{2+})$ remains however too small
  to have any significant effect on derived abundances. 
\item Multiple ionising sources of different temperatures at central 
or non-central locations are not the cause of significant temperature 
fluctuations in the ionised gas of our models with $0.05~\leq~Z/Z_{\odot}~\leq~2$.
\item The relation $t^2$(O$^{+}$) = $t^2$(O$^{2+}$), often used in
  empirical studies, is NOT verified by our models. Extreme care should be taken to account
  for the uncertainties introduced by the use of this relation in
  studies seeking to apply corrections to CEL-derived
  abundances making use of an empirical estimation of temperature
  fluctuations. For our $Z/Z_{\odot}$~=~2 models,
$t^2$(O$^{2+}$) is always a factor of 2 or more higher than
$t^2$(O$^{+}$), while for lower metallicities $t^2$(O$^{2+}$) becomes
lower than $t^2$(O$^{+}$) sometimes by large factors (up to
approximately 10).
\item For intermediate to high metallicity models, electron
  temperatures in the O$^{2+}$ and O$^+$ ionisation zones
  are shifted in opposite directions, contributing to the scatter
  observed in the $T_e(O^{2+})$ versus $T_e(O^{+})$ relation. 
  We confirm that H~{\sc ii} region abundances derived on the basis of the 
T$_e$(O~{\sc ii}) alone should, therefore, be considered highly 
uncertain.
\item Metallicity indicators calibrated by grids of spherically
  symmetric photoionisation models may suffer a systematic bias, due to their
  dependence on the ionisation parameter of the system. For the same
  input parameters case F models will always result in smaller
  ionisation parameters than case C and H models. The errors estimated
  in this work (typically 0.3~dex, but larger in some cases) are
  likely to represent the worst-case scenario, but
  nevertheless their magnitude and their systematic nature does not allow them to be ignored.
\end{enumerate}

{\bf Acknowledgments: }
BE would like to thank the organisers, Fabio Bresolin and Lisa Kewley,
and all the participants to the Metals07 workshop on Metallicity
Calibrations for Gaseous Nebulae (held at the Institute for Astronomy,
of the University of Hawaii on Jan 22-26 2007), for productive
discussion and advice for this work. We would also like to thank Rob
Kennicutt for his many suggestions and Christophe Morisset and Luc
Jamet for their comments and for their careful study of our results. 
BE was partially supported by {\it Chandra} grants GO6-7008X and
GO6-7009X. NB was supported by a PPARC Postdoctoral Fellowship. NB
gratefully acknowledges the hospitality of the Harvard-Smithsonian
Center for Astrophysics, where a significant part of this work took place.
The authors wish to thank the referee Jorge Garc\'ia-Rojas for helpful
comments and constructive discussion.

\end{document}